\documentclass[journal]{IEEEtran}
\usepackage{bm,cite,color,graphicx,epstopdf,authblk,pdfpages,afterpage}
\usepackage[switch]{lineno}
\usepackage[colorlinks=true, allcolors=blue]{hyperref}
\usepackage{hyperref}
\usepackage{mathtools}
\usepackage{multirow}
\usepackage{soul}
\usepackage{subcaption}
\DeclarePairedDelimiter\ceil{\lceil}{\rceil}
\DeclarePairedDelimiter\floor{\lfloor}{\rfloor}

\begin{document}
\title{Knowledge-Based Prediction of Network Controllability Robustness}

\author{Yang~Lou,~\IEEEmembership{}
		Yaodong~He,~\IEEEmembership{}
		Lin~Wang,~\IEEEmembership{Senior Member,~IEEE,}
		Kim~Fung~Tsang,~\IEEEmembership{Senior Member,~IEEE,}
		and~Guanrong~Chen,~\IEEEmembership{Life Fellow,~IEEE}%
\thanks{Y. Lou, Y. He, K.F. Tsang and G. Chen are with the Department of Electrical Engineering, City University of Hong Kong.}%
\thanks{L. Wang is with the Department of Automation, Shanghai Jiao Tong University and also with the Key Laboratory of System Control and Information Processing.}%
\thanks{Supported by NSFC 62002249 and 61873167, and HK ITF ITP/058/17LP.}
\thanks{(\textit{Corresponding author: Guanrong Chen.})}
\thanks{\color{blue}{This paper has been published in \textit{IEEE Transactions on Neural Networks and Learning Systems}, \url{https://doi.org/10.1109/TNNLS.2021.3071367}}.}
}%

\markboth{\url{https://doi.org/10.1109/TNNLS.2021.3071367}~~(April 2021)}%
{Lou \MakeLowercase{\textit{et al.}}: Knowledge-Based Prediction of Network Controllability Robustness}

\maketitle

\begin{abstract}
	Network controllability robustness reflects how well a networked system can maintain its controllability against destructive attacks. Its measure is quantified by a sequence of values that record the remaining controllability of the network after a sequence of node-removal or edge-removal attacks. Traditionally, the controllability robustness is determined by attack simulations, which is computationally time consuming or even infeasible. In the present paper, an improved method for predicting the network controllability robustness is developed based on machine learning using a group of convolutional neural networks (CNNs). In this scheme, a number of training data generated by simulations are used to train the group of CNNs for classification and prediction, respectively. Extensive experimental studies are carried out, which demonstrate that 1) the proposed method predicts more precisely than the classical single-CNN predictor; 2) the proposed CNN-based predictor provides a better predictive measure than the traditional spectral measures and network heterogeneity.
\end{abstract}

\begin{IEEEkeywords}
Complex network, convolutional neural network, controllability, robustness, knowledge-based prediction.
\end{IEEEkeywords}

\IEEEpeerreviewmaketitle

\section{Introduction}
\label{sec:intro}

\IEEEPARstart{C}{omplex} networks as an interdisciplinary research field has gained growing popularity since the late 1990s, encompassing network science, systems engineering, applied mathematics, statistical physics, and biological as well as social sciences \cite{Barabasi2016NS,Newman2010N,Chen2014Book,Chen2019Book}. Scientific studies are trying to understand the essence and characteristics of complex networks while engineering studies are trying to control them for beneficial applications. In the pursuit of networked systems control, whether or not they can be controlled is a fundamental issue, which leads to the basic concept of network controllability  \cite{Liu2011N,Yuan2013NC,Posfai2013SR,Menichetti2014PRL,Motter15CHAOS,Wang2016AUTO,Liu2016RMP,Wang2017RSPTA,Wang2017SR,Zhang2017TAC,Xiang2019CSM}.

The concept of \textit{controllability} refers to the ability of a networked system in changing from any initial state to any desired state under a feasible control input in finite time \cite{Xiang2019CSM}. In retrospect, it was shown that identifying the minimum number of external control inputs (recalled driver nodes), needed to achieve the \textit{structural controllability} of a \textit{directed} network, which requires searching for a maximum matching of the network \cite{Liu2011N}. Thereafter, in \cite{Yuan2013NC}, an efficient measure is introduced for assessing the \textit{state controllability}, based on the rank of the network controllability matrix, for both \textit{directed} and \textit{undirected} networks.

It took quite a long time for people to understand the intrinsic relation between topology and controllability of a general, mostly directed network. It was found that clustering and modularity have no prominent impact on the network controllability, but degree correlation has a certain effect \cite{Posfai2013SR}. It was revealed \cite{Menichetti2014PRL} that random networks of any topology are controllable by an extremely small number of driver nodes if both of its minimum in- and out-degrees are greater than two. A control centrality was introduced in \cite{Liu2012PO} to measure the importance of nodes regarding their roles against random attacks. The network controllability of some canonical graph models is studied and compared in \cite{Wu2018JNS}. For growing networks, the evolution of network controllability is investigated in \cite{Zhang2019PA}. Moreover, the controllability of multi-input/multi-output networked systems is studied in \cite{Wang2016AUTO,Hao2018IJRNC}, with necessary and sufficient conditions derived. Recently, it was realized that some special motifs such as loops and chains are beneficial for enhancing the robustness of network controllability against attacks \cite{Lou2018TCASI,Chen2019TCASII,Lou2019R}. A comprehensive survey of the subject is presented in \cite{Xiang2019CSM}.

Regarding the controllability robustness against attacks, which includes random failures and malicious destructions, a large number of studies have been reported \cite{Holme2002PRE,Shargel2003PRL,Schneider2011PNAS,Liu2012PO,Bashan2013NP,Xiao2014CPB,Wang2018TNSE}. For node- or edge-removal attacks, the main issue is to develop a measure that reflects how well the networks can maintain their controllability after the attacks took place. One measure for the network controllability is quantified by the number of driver nodes needed to recover or retain the network controllability after the occurrence of an attack, while its robustness is quantified by a sequence of values that record the remaining levels of the network controllability after a sequence of attacks \cite{Chen2019TCASII}. To optimize the network robustness, one usually aims to enhance and maintain a highest possible \textit{connectedness} of the network against attacks \cite{Schneider2011PNAS}. Given a degree-preserving constraint (i.e., the degree of each node remains unchanged through the process of optimization), an edge-rewiring method is proposed in \cite{Liang2015CPL}, which increases the number of edges between high-degree nodes, so as to generate a new network with a largest $k$-shell component. In \cite{Chan2016DMKD}, the structure of a network is modified by degree-preserving edge-rewiring, where a spectral measure is employed. By optimizing a specified spectral measure of the network through random edge-rewiring, the robustness of the resultant network is enhanced consequently. However, it was noted that the correlation between spectral measures and the robustness is indeed unclear \cite{Yamashita2019COMPSAC}. Nevertheless, given a reliable predictive measure or indicator of the network robustness, optimization algorithms can be applied \cite{Hou2013ISDEA,Xu2014CCDC,Liu2019ECCN,Wang2019IS}. In the case that there are more than one predictive measure, multi-objective optimization schemes can be adopted \cite{Gunasekara2018MOO,Wang2018TNSE}. In \cite{Zeng2012PRE}, it is shown that network robustness against edge- and node-removals can be enhanced simultaneously. A common observation is that heterogeneous networks with onion-like structures are robust against attacks \cite{Schneider2011PNAS,Wu2011PRE,Tanizawa2012PRE,Hayashi2018SR}. The evolution of alternative attack and defense is studied in \cite{Ma2016PA}, where attack refers to edge-removal and defense means edge-replenishment. The connectedness of the largest-sized cluster is a commonly-used measure for such robustness \cite{Schneider2011PNAS}. It is noted that, although the connectedness robustness has a certain positive correlation with the controllability robustness, they have very different characteristics and measures.

Although the correlation between network topology and network controllability has been investigated, no specific theoretical indicator or performance index was found that can precisely quantify the general network controllability robustness. The nature of the attack methods leads to different measures of the \textit{importance} of a node (or an edge) in a network. In the literature, degree and betweenness are two commonly-used measures for the importance of nodes and edges, respectively \cite{Pu2012PA}.

It was observed that a power-law degree distribution does not necessarily imply a fragile controllability robustness against targeted node-removals; what really contributes to enhance the network controllability robustness is the multi-chain structure \cite{Yan2016SR} and multi-loop structure \cite{Lou2018TCASI,Chen2019TCASII}. Later, it was observed \cite{Lou2019R} that it is particularly beneficial to the network controllability robustness if the multi-loops are across the entire networks rather than only within local communities. Lately it was empirically observed \cite{Lou2020TCASI} that to achieve optimal controllability robustness against random node attacks, both in- and out-degree distributions of a directed network should be extremely homogeneous.

On the other hand, in the field of machine learning, deep neural networks have shown powerful capability in performing classification and regression tasks in image processing. Convolutional neural network (CNN) is one kind of effective deep neural network \cite{Schmidhuber2015NN}. CNN is able to automatically analyze inner features of a dataset without human interference. But, if the user has some prior knowledge and it can be ensured that such prior knowledge would not mislead machine learning, then CNN will become even much powerful for data analysis and processing. Successful real-world applications of CNNs include text recognition and classification \cite{Wang2012ICPR,Lai2015AAAI, Zhang2015NIPS}, face recognition and detection \cite{Li2015CVPR}, image segmentation \cite{Ronneberger2015MICCAI}, etc.

\begin{figure*}[htbp]
	\begin{subfigure}{.25\textwidth}
		\centering
		\includegraphics[width=.7\linewidth]{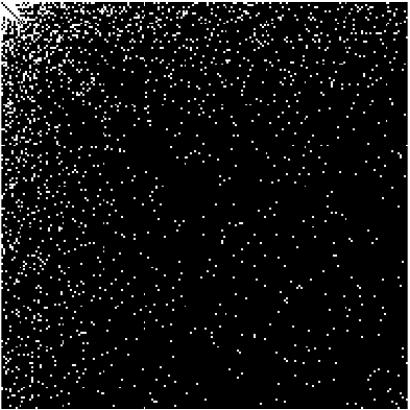}
		\caption{BA unshuffled}
		\label{fig:fig_Ier}
	\end{subfigure}%
	\begin{subfigure}{.25\textwidth}
		\centering
		\includegraphics[width=.7\linewidth]{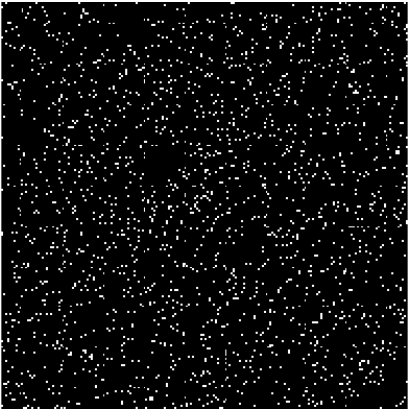}
		\caption{ER unshuffled}
		\label{fig:fig_Isf}
	\end{subfigure}%
	\begin{subfigure}{.25\textwidth}
		\centering
		\includegraphics[width=.7\linewidth]{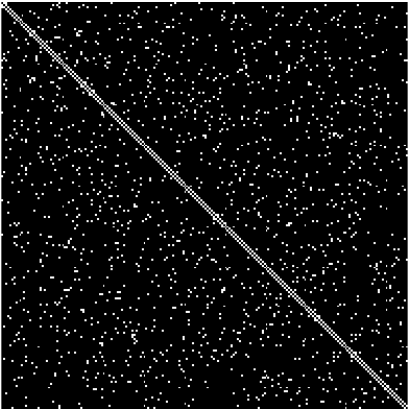}
		\caption{QSN unshuffled}
		\label{fig:fig_Iqs}
	\end{subfigure}%
	\begin{subfigure}{.25\textwidth}
		\centering
		\includegraphics[width=.7\linewidth]{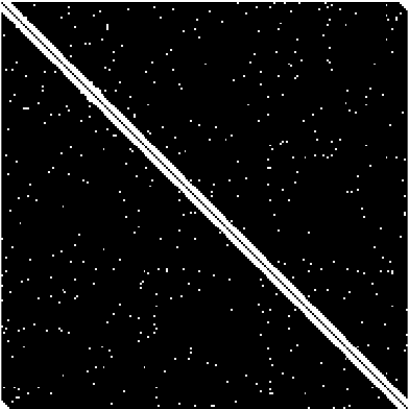}
		\caption{SW unshuffled}
		\label{fig:fig_Isw}
	\end{subfigure}
	\begin{subfigure}{.25\textwidth}
		\centering
		\includegraphics[width=.7\linewidth]{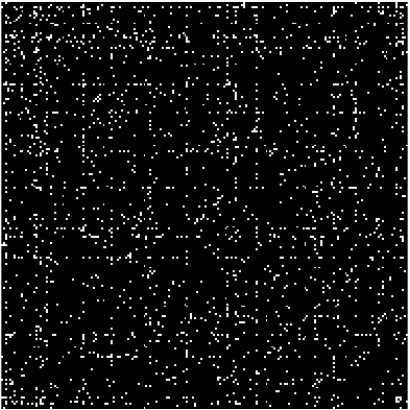}
		\caption{BA shuffled}
		\label{fig:fig_IRer}
	\end{subfigure}%
	\begin{subfigure}{.25\textwidth}
		\centering
		\includegraphics[width=.7\linewidth]{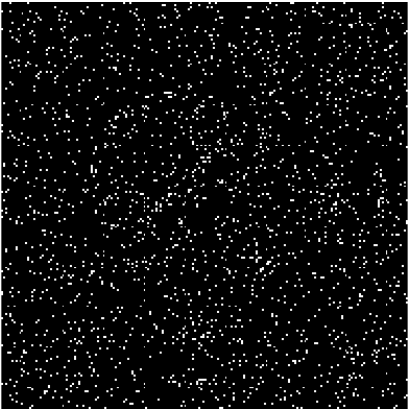}
		\caption{ER shuffled}
		\label{fig:fig_IRsf}
	\end{subfigure}%
	\begin{subfigure}{.25\textwidth}
		\centering
		\includegraphics[width=.7\linewidth]{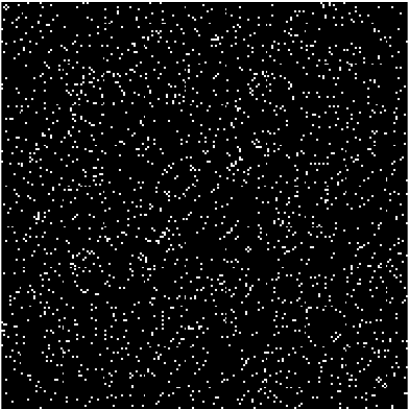}
		\caption{QSN shuffled}
		\label{fig:fig_IRqs}
	\end{subfigure}%
	\begin{subfigure}{.25\textwidth}
		\centering
		\includegraphics[width=.7\linewidth]{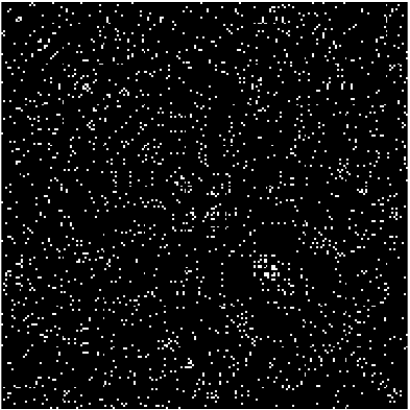}
		\caption{SW shuffled}
		\label{fig:fig_IRsw}
	\end{subfigure}
	\caption{An example of filtering out the generation-based features. The network size $N=200$ with average degree $\langle k\rangle=5.12$.}
	\label{fig:fig_im}
\end{figure*}

Traditionally, for large-scale complex networks, their controllability robustness is evaluated by attack simulations, which however are extremely computationally time consuming. The major computational cost includes: 1) to search for the node to attack, e.g., the maximum betweenness node; 2) the calculation of controllability, e.g., Eq. (\ref{eq:ec}). Both have to be calculated iteratively.
To improve the efficiency of prediction for controllability robustness, this paper takes a machine learning-based approach to designing a knowledge-based predictor for the controllability robustness (iPCR), which is an improved version of the single CNN-based predictor of the controllability robustness (PCR) developed in \cite{Lou2020TCYB}, taking advantage of available prior knowledge.

One unique feature of this iPCR is that it can be applied to both directed and undirected networks, since there is no essential difference for the CNN to process an image converted from a directed or an undirected network, where the symmetry in the graph-converted image does not affect the learning of the CNN. As such, the proposed iPCR has a much wider application range than other traditional methods.

Another improved mechanism in iPCR is that the graph-converted images are updated independently of the generation process. Given an unweighted network, its adjacency matrix can be converted to a black-white image, where a black pixel represents a $0$ element and a white pixel represents a $1$. Given a weighted network with real-valued elements in the adjacency matrix, a gray-scale image is plotted. As shown in Fig. \ref{fig:fig_im}, the upper row shows the intrinsic features of the images pertaining to the generation process. These biased features are filtered out by shuffling the rows and columns of the image, as shown in the lower row of the figure. For example, in a Barab{\'a}si--Albert (BA) scale-free network, the preferential attachment mechanism gives the `old' nodes higher degrees, which are usually allocated near each other therefore have small numbers in the adjacency matrix. As a result, there are always some sparks in the BA-converted image, as shown in Fig. \ref{fig:fig_im} (a). The generation-based features necessarily mitigate the task of a CNN in classification and regression. Therefore, in the iPCR, these special features are filtered out by shuffling the rows and columns of the images, as shown in the lower row of Fig. \ref{fig:fig_im}.

To briefly summarize, the proposed design of the iPCR is based on the following observations: 1) there is no clear correlation between the topological features and the controllability robustness of a general network, directed or undirected, 2) the adjacency matrix of a network can be equivalently represented as a gray-scale image, 3) the CNN technique has proved successful in image processing, and 4) prior knowledge at hand could be sufficiently utilized as preprocessing and filtering tools. In the iPCR, a number of training data generated by simulations are used to train the group of CNNs for classification and prediction, respectively.

Extensive experimental studies are carried out, which demonstrate that 1) the proposed method predicts more precisely than the single-CNN predictor; 2) the CNN-based prediction method provides a better predictive measure than the traditional spectral measures and network heterogeneity.

The rest of the paper is organized as follows: Section \ref{sec:robust} reviews the network controllability and its robustness against various destructive attacks. Section \ref{sec:cnn} introduces the proposed iPCR. In Section \ref{sec:exp}, experimental study is performed with analysis. Finally, Section \ref{sec:end} concludes the investigation.

\section{Network Controllability and Its Robustness}
\label{sec:robust}

Consider a linear time-invariant networked system described by $\dot{{\bf x}}=A{\bf x}+B{\bf u}$, where $A$ and $B$ are constant matrices of compatible dimensions, $\bf x$ is the state vector, and $\bf u$ is the control input. The system is \textit{state controllable} if and only if the controllability matrix $[B\ AB\ A^2B\ \cdots A^{N-1}B]$ has a full row-rank, where $N$ is the dimension of $A$, also the size of the networked system. If a system is state controllable, then its state vector $\bf x$ can be driven from any initial state to any desired state in the state space by a suitable control input $\bf u$ within finite time. The concept of structural controllability is a slight generalization dealing with two parameterized matrices $A$ and $B$, in which the parameters characterize the structure of the underlying networked system: if there are specific parameter values that can ensure the parametric system be state controllable, then the system is \textit{structurally controllable}.

The controllability of a network, or networked system, is measured by the density of the controlled nodes, $n_D$, defined by
\begin{equation}\label{eq:nd}
	n_D\equiv \frac{N_D}{N},
\end{equation}
where $N_D$ is the number of driver nodes needed to retain the network controllability, and $N$ is the network size. This measure $n_D$ allows networks with different sizes can be compared. In comparison, the smaller the $n_D$ value is, the better the network controllability will be.

For a directed network, the number $N_D$ can be calculated according to the \textit{minimum inputs theorem} derived based on maximum matching \cite{Liu2011N}. A maximum matching is a matching that contains the largest possible number of edges, which cannot be further extended in the network. A node is matched if it is the end of an edge in the matching; otherwise, it is unmatched. When a maximum matching is found, the number $N_D$ of driver nodes is determined by the number of unmatched nodes, i.e. $N_D=\text{max}\{1, N-|E^*|\}$, where $|E^*|$ is the number of edges in the maximum matching $E^*$.

As for an undirected network, its controllability can be calculated according to the \textit{exact controllability theorem} derived based on the controllability matrix \cite{Yuan2013NC}. Given an undirected network, its number $N_D$ of driver nodes is calculated by
\begin{equation}\label{eq:ec}
	N_D=\text{max}\{1, N-\text{rank}(A)\}.
\end{equation}

The measure of controllability robustness is calculated by
\begin{equation}\label{eq:ndi}
	n_D(i)\equiv \frac{N_D(i)}{N-i}\,,\ \ i=1,2,\ldots,N-1,
\end{equation}
where $N_D(i)$ is the number of driver nodes needed to retain the network controllability after a total of $i$ nodes have been removed, and $N$ is the original network size. When these values are plotted, a curve is obtained, called the \textit{controllability curve}.

To compare the controllability robustness of two networks against the same attack sequence, their controllability curves are plotted against each other for better visualization. Numerically, a controllability curve $c$ is given by an $(N-1)$ vector $n_D^{c}=[n_D^{c}(1),n_D^{c}(2),\cdots,n_D^{c}(N-1)]$. Thus, given two controllability curves, $c_1$ and $c_2$, the difference (error) of the two curves, when the same number of $i$ nodes are removed, is calculated by
\begin{equation}\label{eq:sigma}
	\sigma(i)=|n_D^{c_1}(i)-n_D^{c_2}(i)|\,.
\end{equation}
The average error $\bar{\sigma}$ is then calculated by
\begin{equation}\label{eq:avg_sigma}
	\bar{\sigma}=\frac{1}{N-1}\sum_{i=1}^{N-1}{\sigma(i)}\,.
\end{equation}
The vector $\sigma(i)$ is used to measure the error between the predicted controllability curve against the true curve; while the scalar $\bar{\sigma}$ measures the overall error of prediction.

The overall network controllability robustness $R_c$ is defined as \cite{Ruths2013CNIV,Xiao2014CPB}
\begin{equation}\label{eq:rc}
	R_c= \frac{1}{N-1} \sum_{i=1}^{N-1}n_D(i)\,,
\end{equation}
where, as an extension of the robustness measure defined in \cite{Schneider2011PNAS}, $n_D(i)$ represents the controllability of the network when a total of $i$ nodes have been removed from the network. Given two complex networks under the same attack, the one with a lower $R_c$ value is considered having better controllability robustness.

In the following, for convenience in description, sometimes the integer index sequence $i=1,2,\ldots,N-1$ will be replaced by the fractional index sequence $p=\frac{1}{N},\frac{2}{N},\ldots,\frac{N-1}{N}$, thereby equivalently replacing $n_D(i)$ ($i=1,2,\ldots,N-1$) with $n_D(p)$ ($p=\frac{1}{N},\frac{2}{N},\ldots,\frac{N-1}{N}$).

\section{Predictor for Network Controllability Robustness}
\label{sec:cnn}

\subsection{Framework of Predictor}
\label{sub:fm}

\begin{figure}[htbp]
	\centering
	\includegraphics[width=.85\linewidth]{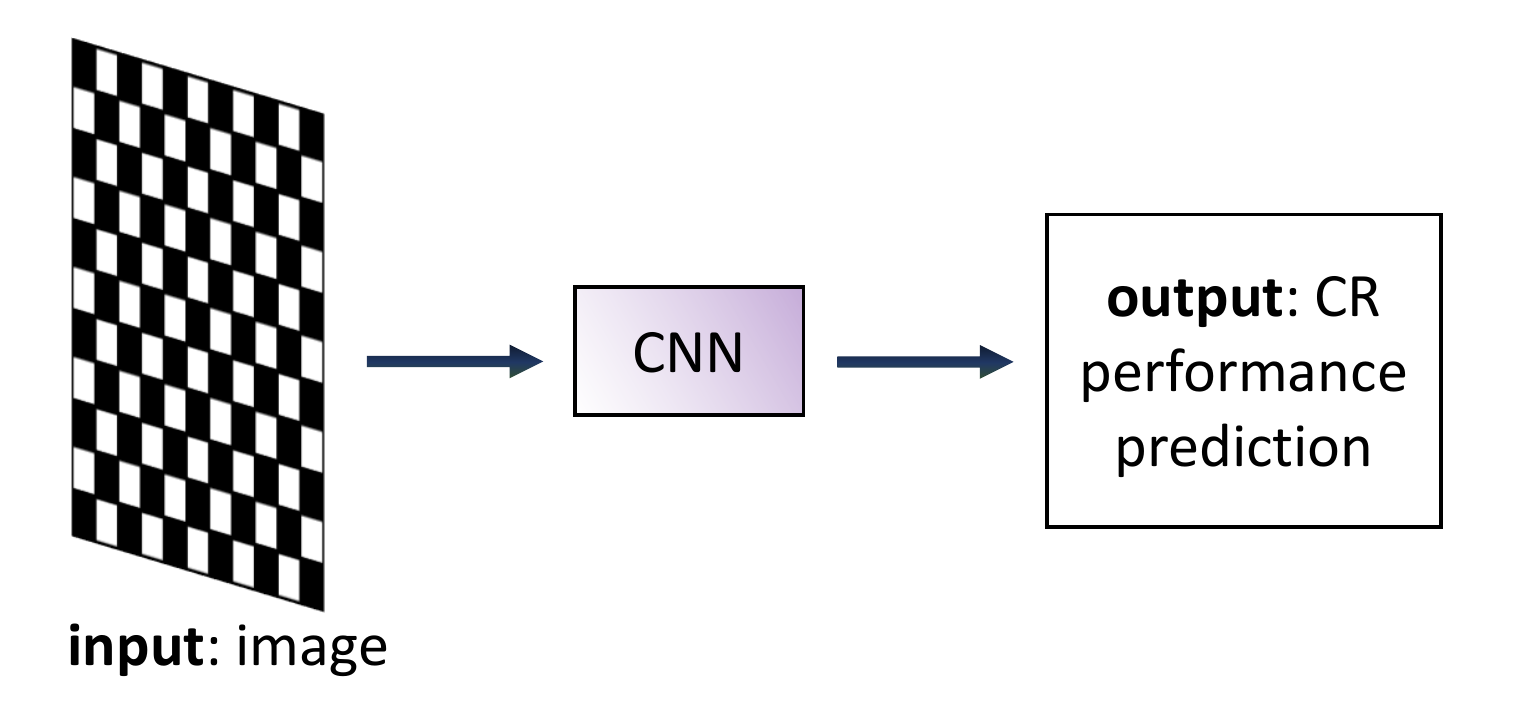}
	\caption{The framework of PCR \cite{Lou2020TCYB}, where a single CNN is used for controllability robustness (CR) prediction. The input is an image converted from the adjacency matrix; the output is the corresponding controllability curve.}
	\label{fig:fc_old}
\end{figure}

The framework of PCR is shown in Fig. \ref{fig:fc_old}, where a single CNN is trained for prediction, referred to as a \textit{predictor}. This framework straightforwardly performs fairly good predictions, with an overall error less than the standard deviation of the testing samples \cite{Lou2020TCYB}. However, there are two main issues about this framework. First, many of the PCR predicted controllability curves are vibrating (especially in the initial stage of the attacks), while the real controllability curves are generally smooth. Second, PCR ignores all available prior knowledge, and trains the single CNN using the raw data without any preprocessing. The proposed iPCR addresses the above two issues by employing a multi-CNN structure, with an extra filter.

To address the first issue, in the new framework a data processer called \textit{filter} is installed after the prediction and before the output. Using available prior knowledge about the dynamics, upper and lower boundaries of the controllability curve can be pre-set. Specifically, during a node-removal attack process, where $i$ ($i=1,2,\ldots,N-1$) nodes are removed, the upper and lower bounds of the controllability curve at position $i$ are pre-set as follows:
\begin{equation}\label{eq:ub}
	ub(i)=\frac{\text{min}(N_D^0+i,N-i)}{N-i}\,,
\end{equation}
and
\begin{equation}\label{eq:lb}
	lb(i)=\frac{1}{N-i}\,,
\end{equation}
where $ub(i)$ and $lb(i)$ represent the upper and lower bounds of $n_D(i)$, respectively; $N_D^0$ means the minimum required number of driver nodes for the original network before being attacked, which can be calculated by Eq. (\ref{eq:ec}).

The following boundary processor is designed:
\begin{equation}\label{eq:bd}
	n_D(i)=\left\{
	\begin{array}{lr}
		ub(i), & \text{if } n_D(i)>ub(i), \\
		lb(i), & \text{if } n_D(i)<lb(i). \\
	\end{array}
	\right.
\end{equation}
Based on this, a median filter with a mask length $L$ is implemented.

To address the second issue, although human-intervention-free is one of the most attractive properties of deep learning, some available knowledge and common sense may be used if such human knowledge would not mislead the machine learning process. Such prior knowledge of the network data will be preprocessed before prediction. For instance, if the network topology is known beforehand, then the prediction work can be passed to a CNN that is specialized for such a topology, which can have better prediction performance.

\begin{figure*}[htbp]
	\centering
	\includegraphics[width=.7\linewidth]{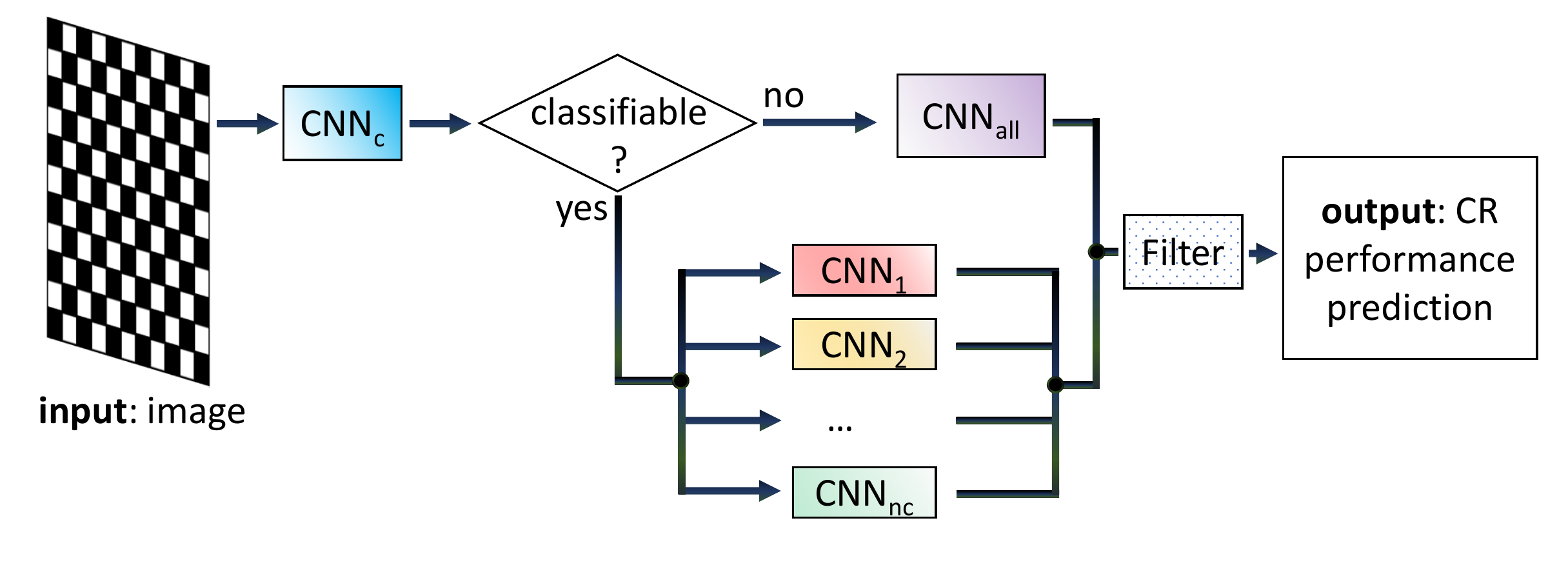}
	\caption{The framework of iPCR, where a CNN\textsubscript{c} is used for network classification. If the input data can be clearly classified as one specific group, then iPCR uses a specifically trained CNN\textsubscript{$i$}, $i=1,2,\ldots,nc$, where $nc$ is the number of clusters; otherwise, if the input data is non-classifiable based on the current knowledge, then iPCR degenerates to PCR using a single CNN\textsubscript{all}.}
	\label{fig:fc_new}
\end{figure*}

iPCR consists of a group of CNNs, including a \textit{classifier} CNN\textsubscript{c} and several \textit{predictors} CNN\textsubscript{$i$} ($i=1,2,\ldots,nc$), as shown in Fig. \ref{fig:fc_new}. All the predictors have the same CNN structure but are trained by different datasets. Each CNN\textsubscript{$i$} ($i=1,2,\ldots,nc$) is trained by the specific cluster of data, such that it is specialized in predicting a cluster, although probably not suitable for another. CNN\textsubscript{all} is trained by all the training data. CNN\textsubscript{c} is trained by applying the prior knowledge of the user.

In the experimental study, two types of prior knowledge are tested, namely the network topology (presented in Sec. \ref{sec:exp}) and the node degree (presented in Supplementary Information (SI)\footnote{\url{https://fylou.github.io/pdf/ipcrsi.pdf}} due to space limitation in the paper). Experimental results show that the former provides helpful prior knowledge, while the later is misleading and consequently the prediction results are degenerated. Finally, before outputting the predicted results, iPCR operates a filter that includes a boundary processor (as shown in Eq. (\ref{eq:bd})) and a median filter.

\subsection{Convolutional Neural Network}
\label{sub:cnn}

The iPCR framework, which includes a classifier, several predictors and a filter, is now introduced along with its configuration and parameter settings.

\begin{figure*}[htbp]
	\centering
	\includegraphics[width=.95\linewidth]{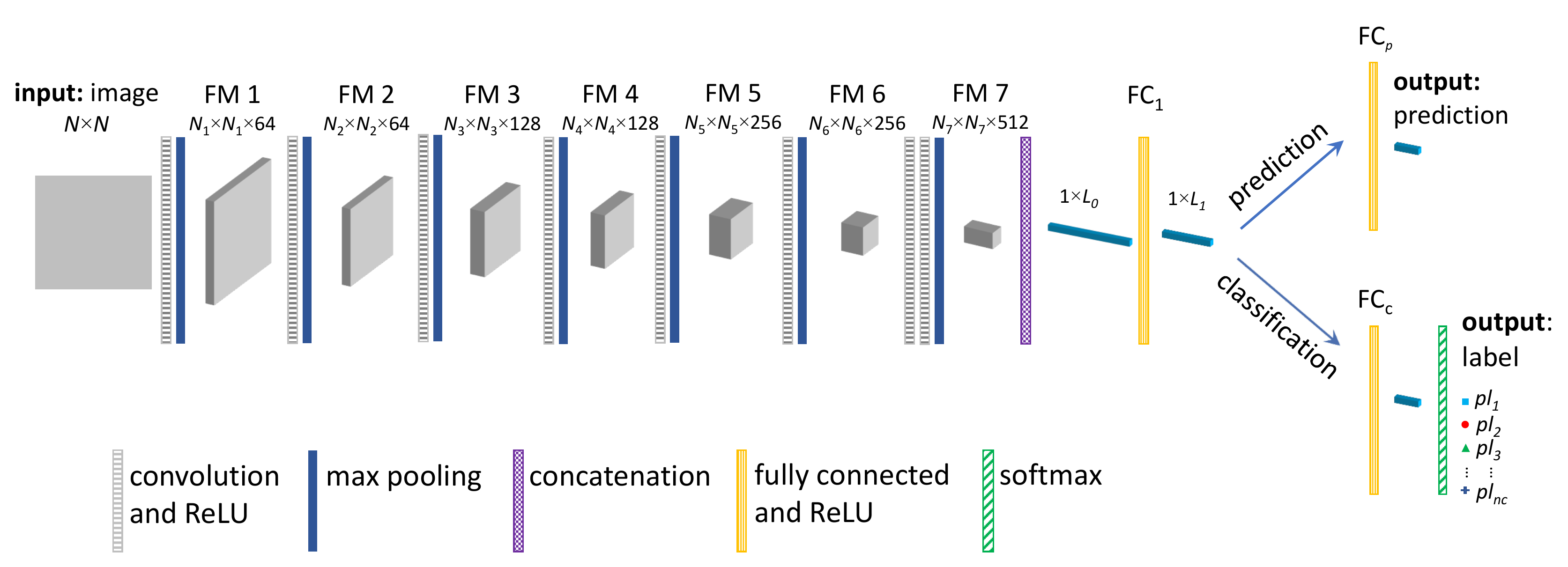}
	\caption{[Color online] The architecture of the CNN used for networks classification and controllability robustness prediction, where FM is an abbreviation for \textit{feature map}, and FC for \textit{fully connected}. The data sizes $N_i=\ceil{N/(i+1)}$, $i=1,2,\ldots,7$. The concatenation layer rearranges the matrix into a vector, from FM $7$ to FC $1$, i.e., $L_0=N_7\times N_7\times 512$. $L_1\in (L_0,N-1)$ is a hyperparameter. Set $L_1=4096$ in this paper. For prediction, another fully-connected layer FC\textsubscript{p} is used as the output layer, yielding an $(N-1)$ vector in the output. For classification, a fully-connected layer FC\textsubscript{c} followed by a softmax layer is used. The output is labeled according to the input data.}
	\label{fig:cnn2}
\end{figure*}

\begin{table}[htbp]
	\centering
	\caption{Parameter settings of the seven groups of convolutional layers.}
	\begin{tabular}{|c|r|c|c|c|} \hline
		Group & \multicolumn{1}{c|}{Layer} & \begin{tabular}[c]{@{}c@{}}Kernel\\ size\end{tabular}
		& Stride & \begin{tabular}[c]{@{}c@{}}Output\\ channel\end{tabular} \\ \hline
		\multirow{2}{*}{Group 1} & Conv7-64 & 7x7 & 1 & 64 \\ \cline{2-5} & Max2 & 2x2 & 2 & 64 \\ \hline
		\multirow{2}{*}{Group 2} & Conv5-64 & 5x5 & 1 & 64 \\ \cline{2-5} & Max2 & 2x2 & 2 & 64 \\ \hline
		\multirow{2}{*}{Group 3} & Conv3-128 & 3x3 & 1 & 128 \\ \cline{2-5} & Max2 & 2x2 & 2 & 128 \\ \hline
		\multirow{2}{*}{Group 4} & Conv3-128 & 3x3 & 1 & 128 \\ \cline{2-5} & Max2 & 2x2 & 2 & 128 \\ \hline
		\multirow{2}{*}{Group 5} & Conv3-256 & 3x3 & 1 & 256 \\ \cline{2-5} & Max2 & 2x2 & 2 & 256 \\ \hline
		\multirow{2}{*}{Group 6} & Conv3-256 & 3x3 & 1 & 256 \\ \cline{2-5} & Max2 & 2x2 & 2 & 256 \\ \hline
		\multirow{2}{*}{Group 7} & Conv3-512 & 3x3 & 1 & 512 \\ \cline{2-5} & Max2 & 2x2 & 2 & 512 \\ \hline
	\end{tabular} \label{tab:cnn_para}
\end{table}

Fig. \ref{fig:cnn2} shows the detailed CNN structure. The detailed parameter settings of the $7$ groups of convolutional layers are given in Table \ref{tab:cnn_para}. Here, the CNN architecture follows the Visual Geometry Group\footnote{\url{http://www.robots.ox.ac.uk/~vgg/}} architecture \cite{Simonyan2014arXiv}. The number of feature map (FM) groups is set to $7$, since the input size is $1000\times1000$ in the following experiments. Note that this number should be set to be greater for a larger input dataset. Each FM consists of a convolution layer, a ReLU, and a max pooling layer. A ReLU provides a commonly-used activation function $f(x)=\text{max}(0,x)$ \cite{Glorot2011ICAIS}. The output of each hidden layer, i.e., a multiplication of weights, is summed up and then rectified by the ReLU for the next layer. The max pooling layer reduces the dimension of the dataset for the input to the next layer.

For prediction, the output is an $(N-1)$ vector that represents the predicted controllability curve; while for classification, the output is a vector of $nc$ real numbers, $pl_i$ ($\sum_{i=1}^{nc}pl_{i}=1$) that represents the probability of the input image belonging to cluster $i$, ($i=1,2,\ldots,nc$). The output layer of the classifier is implemented by adding an extra softmax \cite{Bishop2006} layer in the end.

A threshold $\eta$ is defined for the classifier such that, only if there exists $pl_i\geq\eta$ ($i=1,2,\ldots,nc$), it returns a result indicating that the input image is classifiable (belonging to cluster $i$); otherwise, the input is recognized as non-classifiable. A too-low threshold will decrease the successful rate, while a too-high threshold may result in many non-classifiable cases. In simulations, $\eta$ is set to $0.8$, which yields a successful rate higher than $0.8333$ in classification (see Tables \ref{tab:cm_exp1} and \ref{tab:cm_exp2}). The results will not be sensitively influenced when $\eta$ is slightly changed.

Note that for different purposes, the internal weights of the CNN will be set differently. Here, for illustration, the structures of predictors and classifiers are plotted together. But a predictor and a classifier do not share any internal weight, and each CNN works independently in iPCR.

The loss function used in the classifier is the cross entropy. Given the predicted and the true probability distribution of an instance, denoted by $pl$ and $tl$ respectively, the cross entropy of this instance is calculated as follows:
\begin{equation}\label{eq:xen}
	H=-\sum_{i=1}^{nc} tl(i)\cdot\text{log}[pl(i)]\,.
\end{equation}

The loss function used in the predictor is equal to the mean-squared error between the predicted controllability curve $pv$ and the true curve $tc$, which is calculated as follows:
\begin{equation}\label{eq:lf}
	\mathcal{L} = \frac{1}{N-1} \sum_{i=1}^{N-1}|| pv(i)-tv(i) ||\,,
\end{equation}
where $||\cdot||$ is the Euclidean norm.

The training process for the classifier and predictor aims at adjusting the weights of CNNs, with the objectives of minimizing the cross entropy in Eq. (\ref{eq:xen}) and the mean-squared error in Eq. (\ref{eq:lf}), respectively.

It is worth mentioning that resizing is commonly used for CNNs to process inputs with different sizes. However, therein the application scenario is different. A row and a column represent a node together with all its connected edges, therefore resizing will change the original topology. Nevertheless, the experimental study presented in Sec. \ref{sub:rwn} shows that, if the information loss is very small, the prediction results are still acceptable.  There are a few works that deal with different network sizes by using the same CNN, with additional assumptions, for example, \cite{Niepert2016ICML}. In this paper, CNNs are used to process raw data of complex networks, without any assumption or knowledge on the network structures, and thus the input size can be fixed.

Source codes of this work are available for the public\footnote{\url{https://fylou.github.io/sourcecode.html}}.

\section{Experimental Studies}
\label{sec:exp}

\subsection{Experimental Settings}
\label{sub:set}

Four typical undirected synthetic networks are adopted for simulation, namely the Barab{\'a}si--Albert (BA) scale-free network \cite{Barabasi1999SCI}, Erd{\"{o}}s--R{\'e}nyi (ER) random-graph network \cite{Erdos1964RG}, \textit{q}-snapback network (QSN) \cite{Lou2018TCASI,Lou2019R}, and Newman--Watts (NW) small-world network \cite{Newman1999PLA}.

In the following subsections, the generation methods and parameter settings of the above four networks are introduced, respectively.

Note that, given the network size $N$ and average degree $\langle k\rangle$, there are $M=\floor{N\cdot\langle k\rangle}$ edges in total. Standard notation $\floor{\cdot}$ and $\ceil{\cdot}$ represent the floor and ceiling functions, respectively.

\subsubsection{Barab{\'a}si--Albert (BA) Scale-Free Network}

A BA network is generated as follows:
\begin{itemize}
	\item Start with $n_0$ fully-connected nodes (i.e., an $n_0$-clique).
	\item For nodes $i$ ($i=n_0+1,\ldots,N$), each of them connects to each of nodes $j$ ($j=1,\ldots i-1$) with a probability of $p_{BA}=\frac{k_j}{\sum_{l}k_l}$, where $k_j$ denotes the degree of node $j$. At each step, there are $e_{BA}$ edges being added preferentially.
\end{itemize}
Set $n_0=\ceil{\langle k\rangle}+1$ and $e_{BA}=\frac{M-\binom{n_0}{2}}{N-n_0}$. To exactly control the number of the generated edges to be $M$, proportionally adding or removing edges can be performed.

\subsubsection{Erd{\"{o}}s--R{\'e}nyi (ER) Random-Graph Network}

An ER network is generated as follows:
\begin{itemize}
	\item Start with $N$ isolated nodes.
	\item Pick up all possible pairs of nodes from the $N$ given nodes, denoted as $i$ and $j$ ($i\neq j$, $i,j=1,2,...,N$), once and once only. Connect each pair of nodes with a probability $p_{ER}\in[0,1]$.
\end{itemize}
Let $p_{RG}=\frac{\langle k\rangle}{N-1}$. To exactly control the number of the generated edges to be $M$, uniformly-randomly adding or removing edges can be performed.

\subsubsection{\textit{q}-Snapback Network}

The \textit{q}-snapback network (QSN) was originally constructed as a directed network \cite{Lou2018TCASI} but is converted to be an undirected one here, with only one layer $r_{QSN}$ for simplicity. It is generated as follows:
\begin{itemize}
	\item Start with a chain of $N$ nodes, where each node $i$ ($i=2,...,N-1$) has two edges connecting to its neighboring nodes $i-1$ and $i+1$.
	\item For each node $i=r_{QSN}+1,\, r_{QSN}+2, \ldots, N$, it connects backward to the previously-appeared nodes $i-l\times r_{QSN}$ ($l=1,2,\ldots,\floor{\frac{i}{r_{QSN}}}$), with the same probability $q\in[0,1]$.
\end{itemize}
The probability parameter $q$ can be calculated from the given $N$, $M$, and $r_{QSN}$. For $r_{QSN}=1$, $q=\frac{(M-N)\cdot{r_{QSN}}}{\sum_{j=2+r_{QSN}}^{N-2}{j}}= \frac{M-N}{\sum_{j=3}^{N-2}{j}}$. To exactly control the number of the generated edges to be $M$, uniformly-randomly adding or removing edges can be performed.

\subsubsection{Newman--Watts (NW) Small-World Networks}

An NW network is generated as follows:
\begin{itemize}
	\item Start with an $N$-node loop having $K$ connected nearest-neighbors on each side.
	\item Some edges are added without removing any existing edges, until totally $M$ edges have been added.
\end{itemize}
Set $K=2$ in the following; that is, a node $i$ is connected to its two nearest neighbors on each side, i.e., with nodes $i-1$, $i+1$, $i-2$ and $i+2$.

Since the above generation methods will generate networks with some strong visible features (as illustrated by Fig. \ref{fig:fig_im}), the rows and columns of the resulting adjacency matrices are shuffled and random isomorphs are generated, so as to filter out these undesirable features.

The training data are generated by performing attack simulations on the generated networks, such that the controllability curves (see Eq. (\ref{eq:ndi})) can be obtained. These graph-converted images and controllability curves are used for iPCR training. Given an input graph-converted image, the trained iPCR can be used to predict its controllability curve, skipping the time-consuming attack simulation process.

Next, the prediction performances of PCR and iPCR are compared on 1) unweighted networks against random node-removal attacks (see Sec. \ref{sub:exp1}), 2) weighted networks against targeted node-removal attacks (see Sec. \ref{sub:exp2}); 3) real-world networks under random attacks (see Sec. \ref{sub:rwn}). In these three experimental studies, PCR and iPCR aim to predict precise controllability curves. In Sec. \ref{sub:cmp}, PCR and iPCR are compared on predicting the ordinal ranks of the network controllability robustness, with respect to $6$ spectral measures and the heterogeneity. Finally, the computational costs are briefly discussed in Sec. \ref{sub:cost}. In all the following comparisons, a filter consisting of a boundary processor and a median filter (with $L=9$) are installed in both PCR and iPCR.

\subsection{Unweighted Networks Under Random Attacks}
\label{sub:exp1}

The controllability robustness prediction on unweighted networks with size $N=1000$ and average degree $\langle k\rangle=3$, $4$, and $5$, under random node-removal attacks, is studied.

There are $12$ network configurations in total. For each configuration, $500$ training samples are used. Each sample includes an adjacency matrix (as the input) and its controllability curve obtained from simulation (as the output). CNN\textsubscript{all} of iPCR is trained by $12\times500=6000$ training samples; while each of CNN\textsubscript{1,2,3,4} is trained by $3\times500=1500$ samples. Each CNN\textsubscript{k} ($\text{k}=1,2,3,4$) is specifically trained for one of the four network types, namely BA, ER, QSN, and NW; while CNN\textsubscript{all} is trained by the ensemble of all the networks. PCR is trained in the same way as CNN\textsubscript{all} of iPCR. Given a random attack, in simulation the result is averaged from $10$ independent runs, so as to balance mitigating and randomness influences, which also reduces the burden of computation.

Another set of $100$ testing samples for each network configuration are generated independently. The classification results of CNN\textsubscript{c} are shown in Table \ref{tab:cm_exp1}. As illustrated by Fig. \ref{fig:fig_im}, shuffling filters out the method-generated features in the resultant images,  resulting images indistinguishable by eyes, which makes the classification task becoming tougher. It can be seen from Table \ref{tab:cm_exp1} that CNN\textsubscript{c} correctly classifies the four types of networks at a successful rate higher than $0.8333$. Since the threshold is set to $\eta=0.8$ in the softmax layer of the classifier, an input is non-classifiable if it generates a result with the probability of success less than $0.8$, and in this case the input will be passed to CNN\textsubscript{all} for prediction. As can be seen from Table \ref{tab:cm_exp1}, the rate of non-classifiable (NC) data is low, indicating the effectiveness of the classifier which uses prior knowledge.

Note that if an input is incorrectly classified, it will be passed to a wrong predictor that is specialized for a different network type. This will totally mislead the prediction, and therefore is harmful. In iPCR, according to Table \ref{tab:cm_exp1}, BA and QSN may be mis-classified as ER at rates $0.0922$ and $0.0165$, respectively; NW may be mis-classified QSN at a rate $0.0042$; ER will not be mis-classified to other networks, but becomes non-classifiable at rate $0.0254$. Overall, the classification error rates are quite low, so iPCR is proved working well.

\begin{table}[htbp]
	\centering
	\caption{Confusion matrix of the CNN\textsubscript{c} classifier for classifying unweighted networks. NC means the input is non-classifiable; (\textit{pre}) represents the predicted type and (\textit{act}) represents the actual type of the network.}
	\begin{tabular}{|c|c|c|c|c|c|}
		\hline
		&\begin{tabular}[c]{@{}c@{}}BA\\(\textit{pre})\end{tabular} & \begin{tabular}[c]{@{}c@{}}ER\\ (\textit{pre})\end{tabular} & \begin{tabular}[c]{@{}c@{}}QSN\\(\textit{pre})\end{tabular} & \begin{tabular}[c]{@{}c@{}}NW\\ (\textit{pre})\end{tabular} & NC \\ \hline
		\begin{tabular}[c]{@{}c@{}}BA\\(\textit{act})\end{tabular}  & 0.8333 & 0.0922 & 0 & 0 & 0.0745 	\\ \hline
		\begin{tabular}[c]{@{}c@{}}ER\\(\textit{act})\end{tabular}  & 0 & 0.9746 & 0 & 0 & 0.0254 		\\ \hline
		\begin{tabular}[c]{@{}c@{}}QSN\\(\textit{act})\end{tabular} & 0 & 0.0165 & 0.9342 & 0 & 0.0494 	\\ \hline
		\begin{tabular}[c]{@{}c@{}}NW\\(\textit{act})\end{tabular}  & 0 & 0 & 0.0042 & 0.9833 & 0.0126 	\\ \hline
	\end{tabular} \label{tab:cm_exp1}
\end{table}

\begin{figure*}[htbp]
	\centering
	\includegraphics[width=.95\linewidth]{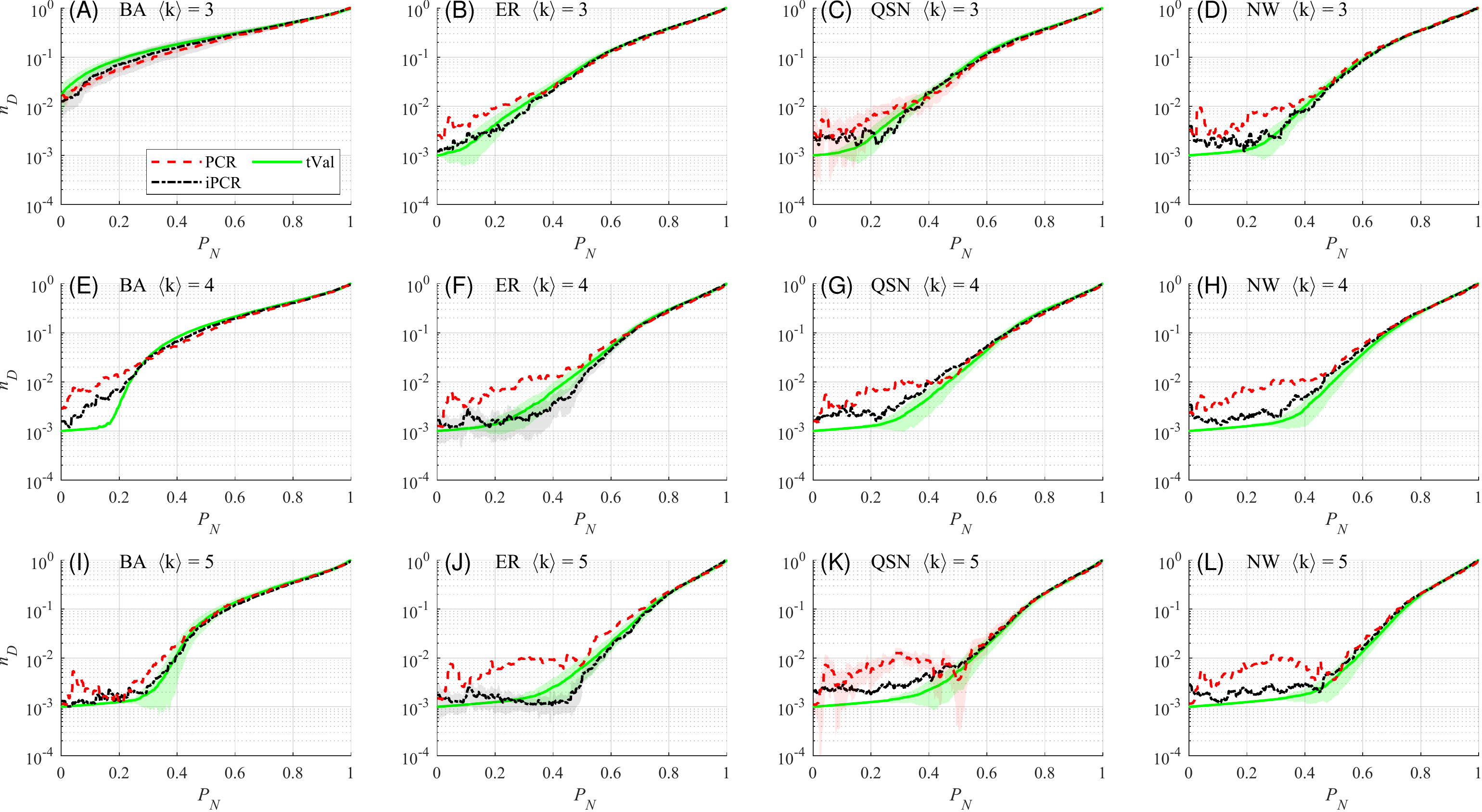}
	\caption{[Color online] Comparison of PCR and iPCR on unweighted networks under random attacks. $P_N$ represents the portion of nodes having been removed from the network; $n_D$ is calculated by Eq. (\ref{eq:nd}). Green curves: the true value (tVal) from simulation; red curves: predicted by PCR; black curves: predicted by iPCR. The shaded shadow in the same color represents the range of standard deviation.}
	\label{fig:exp1}
\end{figure*}

The performance comparison between PCR and iPCR is shown in Fig. \ref{fig:exp1}. In each subplot, in a unique network configuration, the green curve shows the true value (tVal) generated by simulation; the red dashed curve shows the predicted values by PCR; the black dotted curve represents the predicted results of iPCR. The shadow in the same color represents the range of standard deviation. As can be seen from the plots, the black curves are obviously closer to the green curves, better than the red curves, meaning that iPCR predicts the controllability more accurately than PCR, in all $12$ cases. The results confirm that prior knowledge of the network topology is indeed helpful if correctly used. It is notable that the predicted curves are not as oscillatory as those obtained in \cite{Lou2020TCYB}, thanks to the filters used in both PCR and iPCR.

\begin{figure}[htbp]
	\begin{subfigure}{.25\textwidth}
		\centering
		\includegraphics[width=.8\linewidth]{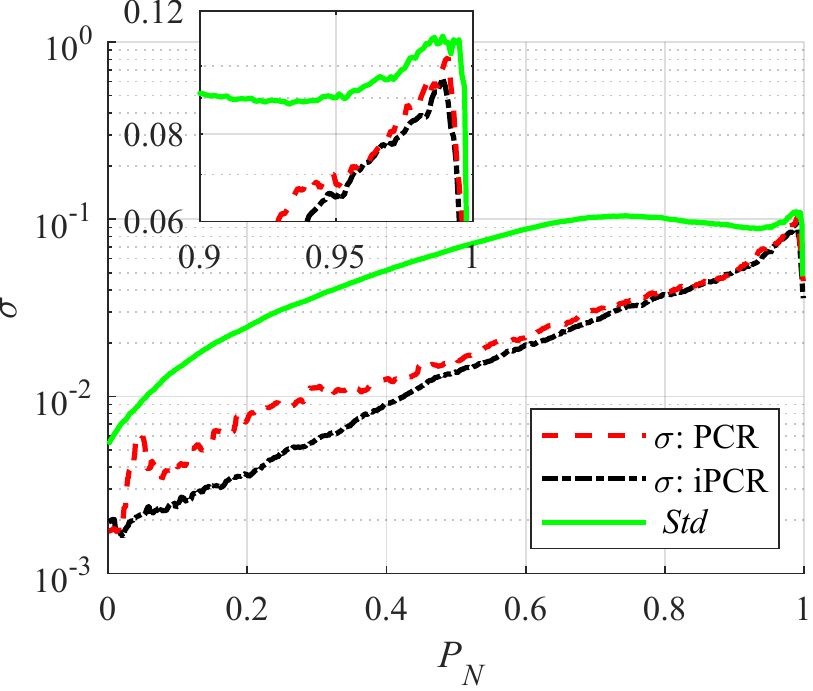}
		\caption{$N$=$1000$, the inset gives a zoom-in for $P_N\in [0.9,1]$.}
	\end{subfigure}%
	\begin{subfigure}{.25\textwidth}
		\centering
		\includegraphics[width=.8\linewidth]{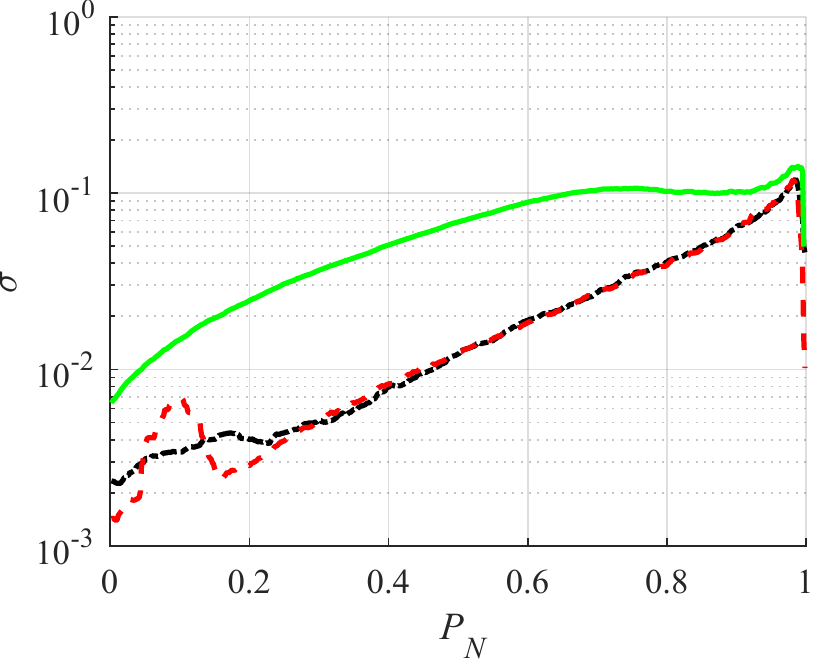}
		\caption{$N$=$600$}
	\end{subfigure}
	\begin{subfigure}{.25\textwidth}
		\centering
		\includegraphics[width=.8\linewidth]{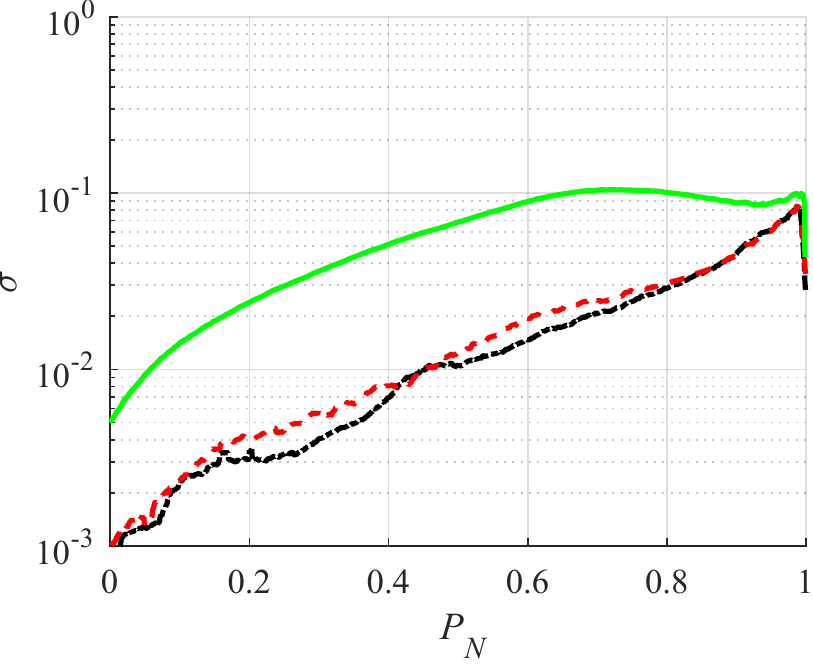}
		\caption{$N$=$1200$}
	\end{subfigure}%
	\begin{subfigure}{.25\textwidth}
		\centering
		\includegraphics[width=.8\linewidth]{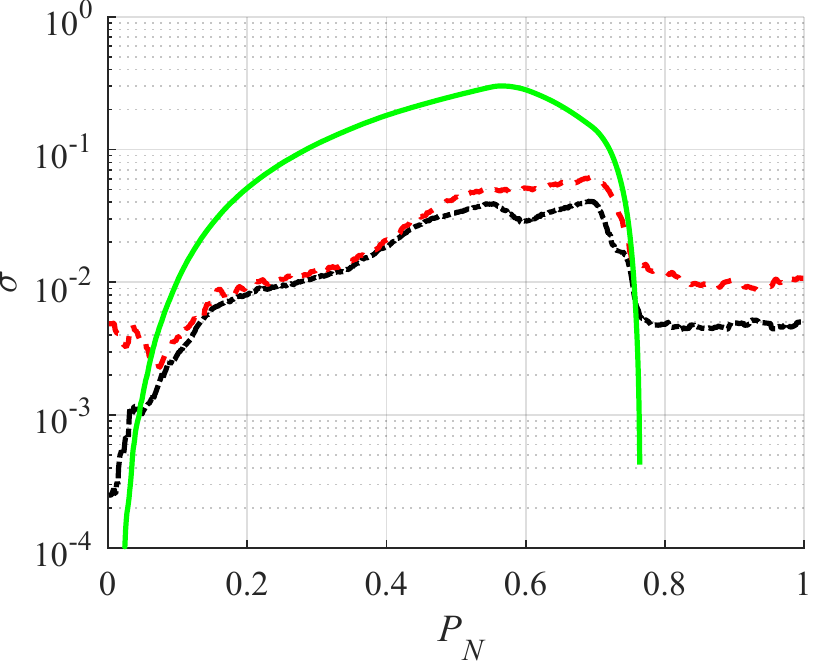}
		\caption{weighted, targeted attacks}
	\end{subfigure}
	\caption{[Color online] The average errors ($\sigma$) comparison of PCR and iPCR: unweighted networks with (a) $N=1000$, (b) $N=600$, and (c) $N=1200$, under random attacks; (d) weighted networks ($N=1000$) under targeted attacks. The green curve (\textit{Std}) represents the standard deviation of the testing samples.}
	\label{fig:err}
\end{figure}

The average error calculated according to Eq. (\ref{eq:sigma}) is plotted in Fig. \ref{fig:err} (a), where the black curve shows that iPCR has a lower average error ($\sigma$) than the red dashed curve of PCR, through the entire attacking process. The inset shows a clearer plot of the comparison for $P_N\in [0.9,1]$. The average error is taken from all $12$ configurations, namely, a total of $1200$ testing samples. It is noticeable that both PCR and iPCR gain average errors with standard deviations much lower than that of the testing set (i.e., the $1200$ testing samples), throughout the entire attack process.

To verify the scaling property, given the same conditions except for the network size, the prediction performances of iPCR and PCR are compared. Figs. \ref{fig:err} (b) and (c) show the average errors ($\sigma$) comparisons when $N=600$ and $1200$, respectively. The detailed predicted controllability curves are shown in SI.

\subsection{Weighted Networks Under Targeted Attacks}
\label{sub:exp2}

The controllability robustness prediction on weighted networks with size $N=1000$ and average degree $\langle k\rangle\in[3,5]$, under targeted node-removal attacks, is studied.

For each network instance, its average degree $\langle k\rangle$ is a real random number generated from the range of $[3,5]$; its edge weights are uniformly-randomly assigned from the range of $(0,1)$. Again, PCR contains a single CNN, while iPCR uses a CNN\textsubscript{all} with four specialized CNN\textsubscript{1,2,3,4} for the four types of networks respectively, if classifiable. The targeted attack performs node-removals according to the degrees of nodes, from high to low sequentially.

\begin{table}[htbp]
	\centering
	\caption{Confusion matrix of the CNN\textsubscript{c} classifier on classifying weighted networks. NC means the input is non-classifiable; (\textit{pre}) represents the predicted type and (\textit{act}) represents the actual type of the network; initial `w' is for `weighted'.}
	\begin{tabular}{|c|c|c|c|c|c|}
		\hline
		&\begin{tabular}[c]{@{}c@{}}wBA\\(pre)\end{tabular} & \begin{tabular}[c]{@{}c@{}}wER\\ (pre)\end{tabular} & \begin{tabular}[c]{@{}c@{}}wQSN\\(pre)\end{tabular} & \begin{tabular}[c]{@{}c@{}}wNW\\ (pre)\end{tabular} & UC     \\ \hline
		\begin{tabular}[c]{@{}c@{}}wBA\\(act)\end{tabular}  & 0.9913 & 0 & 0 & 0 & 0.0087		\\ \hline
		\begin{tabular}[c]{@{}c@{}}wER\\(act)\end{tabular}  & 0 & 0.9549 & 0.0150 & 0 & 0.0301	\\ \hline
		\begin{tabular}[c]{@{}c@{}}wQSN\\(act)\end{tabular} & 0 & 0 & 0.9915 & 0 & 0.0085		\\ \hline
		\begin{tabular}[c]{@{}c@{}}wNW\\(act)\end{tabular}  & 0 & 0 & 0.0074 & 0.9815 & 0.0111 \\ \hline
	\end{tabular}  \label{tab:cm_exp2}
\end{table}

The confusion matrix shown in Table \ref{tab:cm_exp2} suggests that the precision of the CNN\textsubscript{c} classifier is high. Slightly different from Table \ref{tab:cm_exp1}, here the CNN\textsubscript{c} can either correctly classify the weighted BA and QSN respectively, or return a result of non-classifiable, without any mis-classification. The weighted ER and NW have very low probabilities to be classified as weighted QSN. Shuffling is also performed on these weighted networks. The overall precision on classifying weighted networks is slightly higher than that on unweighted networks.

In the experiments reported in Sec. \ref{sub:exp1}, the average degree $\langle k\rangle$ is set to integers $3,4,5$, respectively. Each column in Fig. \ref{fig:exp1} shows the same type of networks, with increasing $\langle k\rangle$ from $3$ to $5$. Although PCR and iPCR are trained without any information about the average degrees, both PCR and iPCR return different predictions, when the input is of the same network type with different average degrees. However, this does not imply that average degree is a good feature or useful prior knowledge. In contrast, the average degree is known to be not suitable for preprocessing when used as prior knowledge. An example is given in the SI, where three network clusters are defined, namely `$\langle k\rangle=3$', `$\langle k\rangle=4$', and `$\langle k\rangle=5$'. The prediction results is distorted due to the low precision of classification. This demonstrates that the prior knowledge used should be correct and appropriate, as common sense, otherwise misleading could happen.

\begin{figure}[htbp]
	\centering
	\includegraphics[width=.9\linewidth]{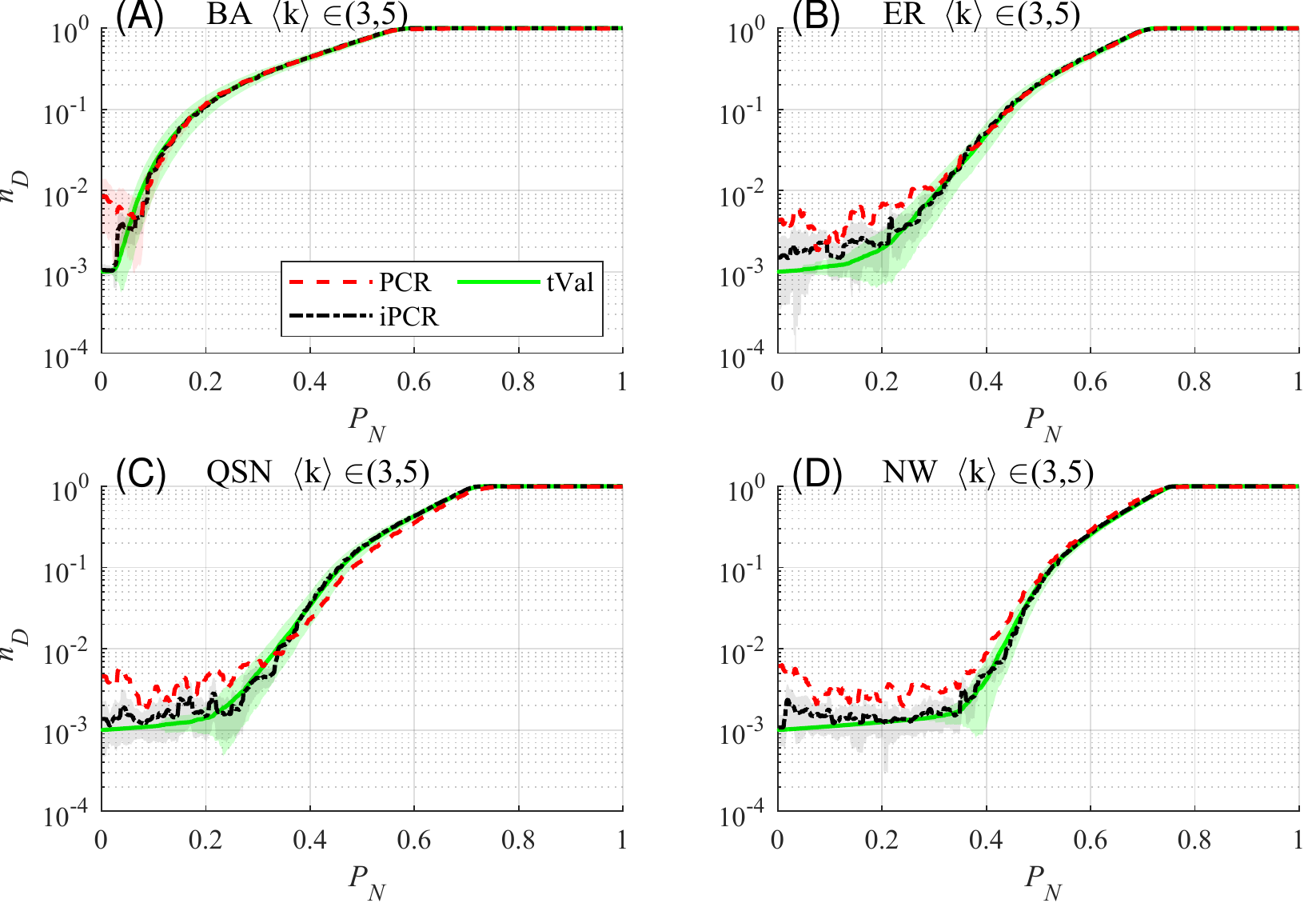}
	\caption{[Color online] Comparison of PCR and iPCR on weighted networks under targeted attacks. $P_N$ represents the portion of nodes having been removed from the network; $n_D$ is calculated by Eq. (\ref{eq:nd}). Green curves: the true value (tVal) from simulation; red curves: predicted by PCR; black curves: predicted by iPCR. The shaded shadow in the same color represents the range of standard deviation.}
	\label{fig:exp2}
\end{figure}

Fig. \ref{fig:exp2} shows the prediction results of PCR and iPCR on weighted networks under targeted attacks. Again, it is clear that, in each subplot, the black dotted curve is closer to the green curve than the red dashed curve. The higher precision of iPCR in prediction is partially due to the high classification rate presented in Table \ref{tab:cm_exp2}. Another reason is that an specialized CNN predictor is always better than a mixed one, as is intuitively so.

Fig. \ref{fig:err} (d) shows that the average prediction error of iPCR (black curve) is lower than that of PCR (red dashed curve), throughout the entire attack process. Note that both PCR and iPCR gain average errors with standard deviations much lower than that of the testing samples through a long period. Differing from random attacks, in a targeted attack, when the portion of removed nodes is somewhat greater than $0.7$, the network requires $n_D\approx 1$ to gain a full controllability. Although PCR and iPCR gain lower predictive errors during this stage (when $P_N$ is somewhat greater than $0.7$), the standard deviation of the testing sample actually becomes nearly zero.

\subsection{Real-world Networks Under Random Attacks}
\label{sub:rwn}

The PCR and iPCR trained in Sec. \ref{sub:exp1} are used for predicting the controllability robustness of $6$ real-world networks with  $N\approx1000$. Basic information of these networks is given in Table \ref{tab:rwn}; the network data are from Network Repository \footnote{\url{http://networkrepository.com/}}. 

\begin{table}[htbp]
	\centering
	\caption{Basic information of the real-world networks.}
	\begin{tabular}{|c|c|l|c|c|}
		\hline
		\begin{tabular}[c]{@{}c@{}}abbr.\\ name\end{tabular} & file name & \multicolumn{1}{c|}{brief description} & $N$ &$M$  \\ \hline
		DDG & DD-g79 & protein & 1022 & 2889 \\ \hline
		DEL & delaunay-n10 & DIMACS10 problem & 1024 & 3056 \\ \hline
		DW5 & dwt-1005 & & 1005 & 3808 \\ \cline{1-2} \cline{4-5} DW7 & dwt-1007 & \multirow{-2}{*}{ \begin{tabular}[c]{@{}l@{}}symmetric connection\\ from Washington\end{tabular}} & 1007 & 3784 \\ \hline
		LSH & lshp1009 & \begin{tabular}[c]{@{}l@{}}Alan George's\\ L-shape problem\end{tabular} & 1009 & 2928 \\ \hline
		ORS & orsirr-1 & \begin{tabular}[c]{@{}l@{}}oil reservoir\\ simulation\end{tabular} & 1030 & 2914 \\ \hline
	\end{tabular}\label{tab:rwn}
\end{table}

\begin{figure}[htbp]
	\begin{subfigure}{.5\textwidth}
		\centering
		\centering
		\includegraphics[width=\linewidth]{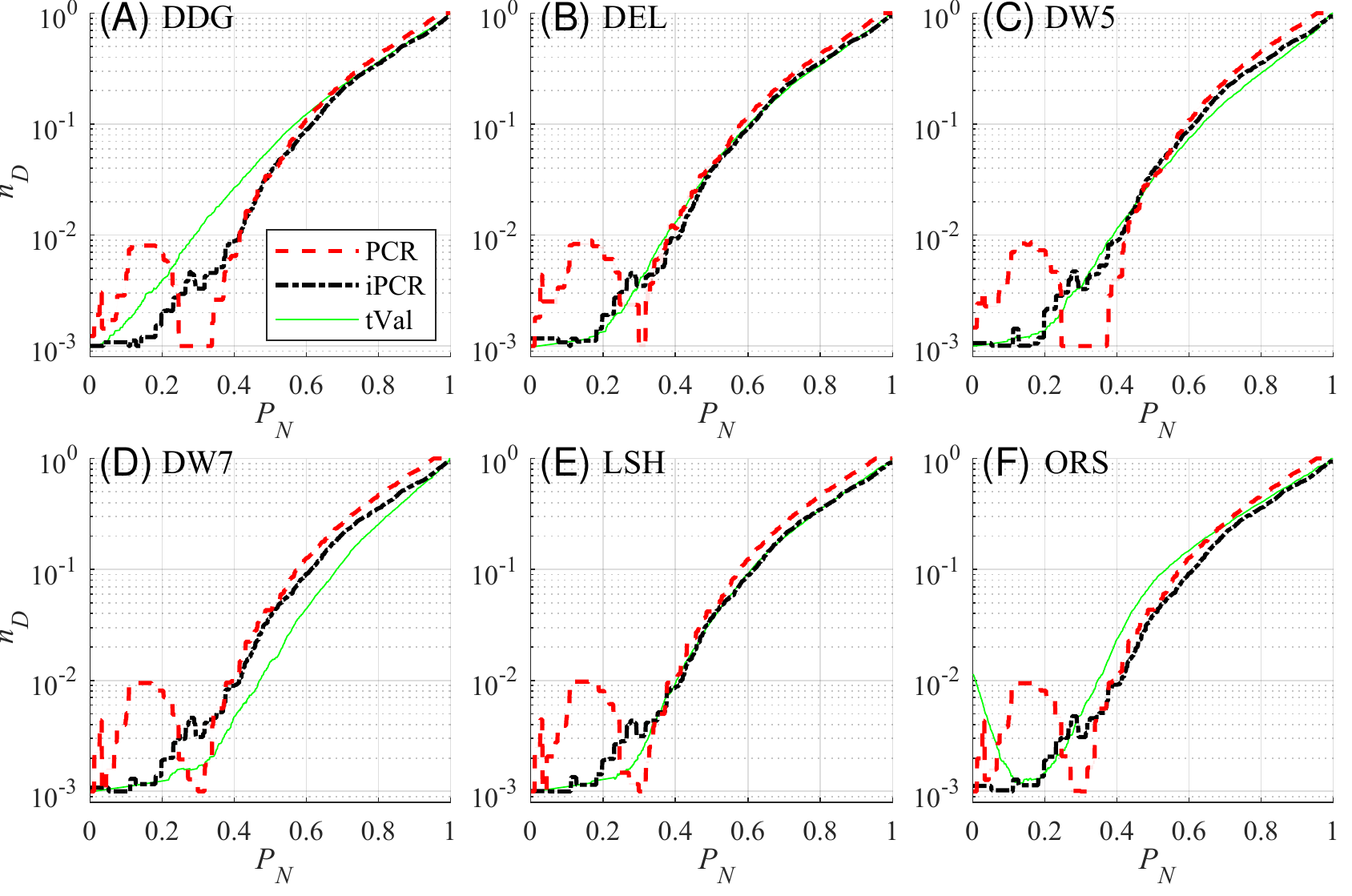}
		\caption{Prediction Results}
	\end{subfigure}
	\begin{subfigure}{.5\textwidth}
		\centering
		\centering
		\includegraphics[width=\linewidth]{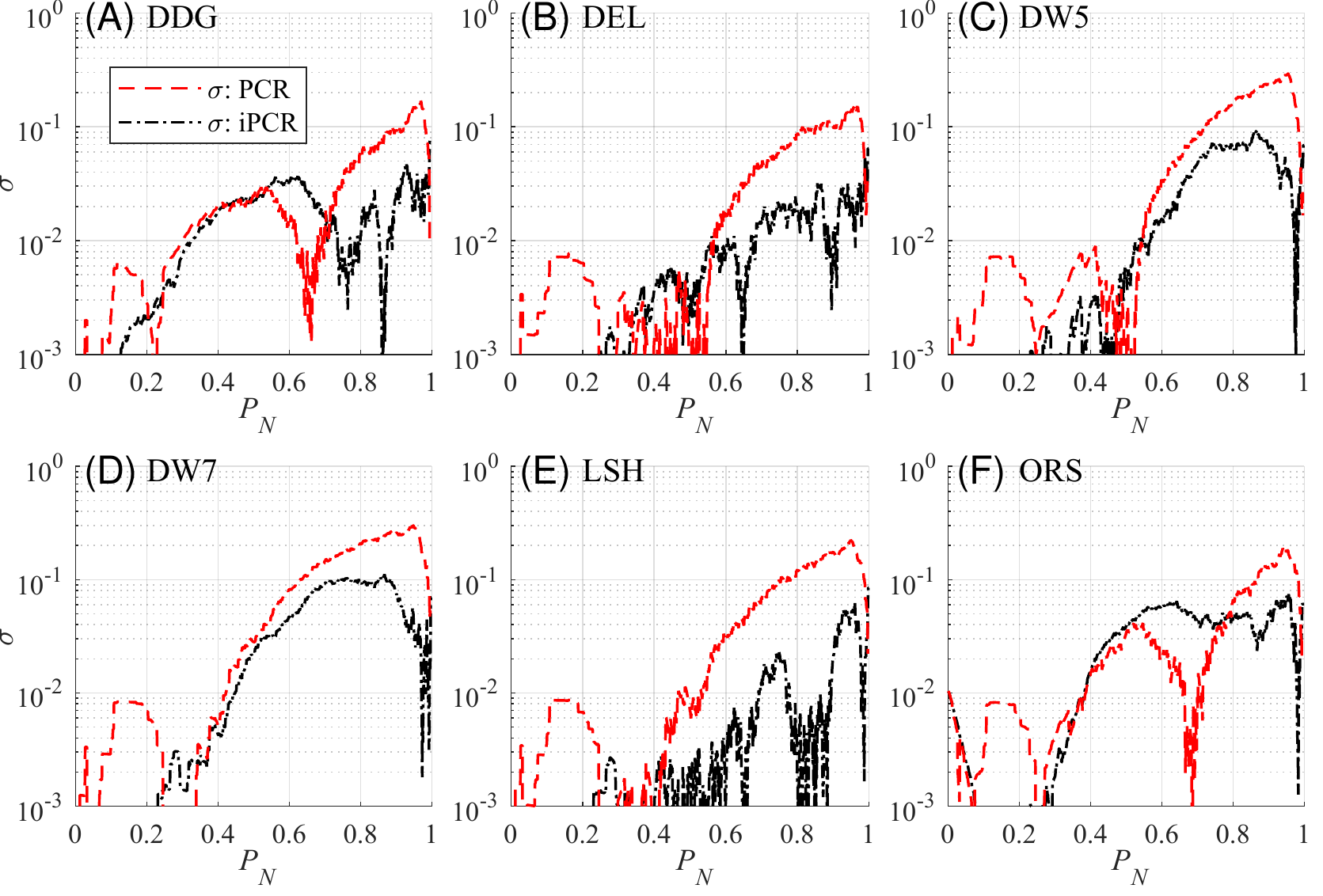}
		\caption{Prediction Errors}
	\end{subfigure}	
	\caption{[Color online] Performance comparison of PCR and iPCR on real-world networks under random attacks: (a) prediction results; (b) prediction errors.}
	\label{fig:exp4}
\end{figure}

Since the sizes of the real-world networks are slightly larger than $1000$ (see Table \ref{tab:rwn}), resizing is performed on the graph-converted images, i.e., a pair of row and column is randomly picked and removed until it reaches $N=1000$. For each network, the random resizing is repeated  for $20$ times, and the prediction results and errors are then averaged. As can be seen from Fig. \ref{fig:exp4} (a), iPCR predicts the controllability curves closer to the true curves than PCR does, especially in the early stage. Fig. \ref{fig:exp4} (b) shows that iPCR obtains clearly lower prediction errors than PCR for networks DEL, DW5, DW7 and LSH; while for DDG and ORS, PCR obtains lower errors than iPCR only in a limited period at the middle stage. Both PCR and iPCR predict the controllability performances of these real-world networks with low errors.

\subsection{Comparison of Prediction Measures}
\label{sub:cmp}

Spectral measures have long been used to quantify the \textit{connectedness} robustness of complex networks against node- and edge-removal attacks. It has certain positive correlation to controllability, but they cannot be treated equally.

Here, six commonly-used spectral measures, namely spectral radius (SR), spectral gap (SG), natural connectivity (NCo), algebraic connectivity (ACo), effective resistance (ERe), and spanning tree count (STC) are compared in measuring the controllability robustness. Definitions and computational formulas for these measures can be found in, e.g., \cite{Chan2016DMKD}. Recently, it was also found that heterogeneity (HE) reflects the controllability robustness \cite{Lou2019R}. In this paper, PCR and iPCR are used to predict the entire controllability curves in Secs. \ref{sub:exp1} and \ref{sub:exp2}. Noticed that a predicted curve (a vector) can also be converted to a measure (a scalar) through Eq. (\ref{eq:rc}). Thus, in the following, the above $9$ prediction measures will be compared, namely, the $6$ common spectral measures (SR, SG, NCo, ACo, ERe, STC), HE, PCR and iPCR.

The above prediction measures are used to predict the ordinal ranks of a total of $1200$ networks, for four network types with three different average degrees. These ranks are listed in a descending order in terms of controllability robustness, from the best to the worst. Specifically, each prediction measure returns a predicted rank list of the $1200$ networks. Then, the $9$ rank lists returned by the $9$ prediction measures are compared to the true rank list generated by simulation. The rank error information is summarized in Table \ref{tab:rank_er}, where the rank error $\sigma_{rank}$ is calculated by
\begin{equation}\label{eq:erank}
	\sigma_{rank}=|rl_{pm}-rl_t|\,,
\end{equation}
where $rl_{pm}$ represents the rank list predicted by a prediction measure, and $rl_t$ represents the true rank obtained from simulation.

For example, given two predicted rank lists, $rl_{pm1}=[3,5,2,4,1]$ and $rl_{pm2}=[2,1,4,5,3]$, and a true rank list, $rl_t=[1,2,3,4,5]$, the rank errors are obtained as $\sigma_{rank1}=[2,3,1,0,4]$, $\sigma_{rank1}=[1,1,1,1,2]$, respectively. The mean, maximum, and minimum of the rank error can be calculated accordingly. The number of `$0$' elements in a rank error list $\sigma_{rank}$ is counted and then included in the `correct rank' column. Finally, the number of networks, which are predicted top $10\%$ and then confirmed top $10\%$ by simulation, in terms of controllability robustness, is counted and included in the `top $10\%$' column. The number in the `bottom $10\%$' column is similarly calculated. The detailed rank values of the $9$ prediction measures are given in SI.

\begin{table}[htbp]
	\centering
	\caption{Rank error information for the $9$ predictive measures. Bold number means the best performing prediction measures.}
	\begin{tabular}{|c|c|c|c|c|c|c|}
		\hline
		& \begin{tabular}[c]{@{}c@{}}average\\ rank\\ error\end{tabular} & \begin{tabular}[c]{@{}c@{}}max\\ rank\\ error\end{tabular} & \begin{tabular}[c]{@{}c@{}}min\\ rank\\ error\end{tabular} & \begin{tabular}[c]{@{}c@{}}correct\\ rank\end{tabular} & \begin{tabular}[c]{@{}c@{}}top\\ 10\%\end{tabular} & \begin{tabular}[c]{@{}c@{}}bottom\\ 10\%\end{tabular} \\ \hline
		SR  & 387.52 & 912  & \textbf{0} & 3 & 0 & 5 \\ \hline
		SG  & 370.09 & 933  & 1 & 0 & 0  & 7 \\ \hline
		NCo & 394.58 & 993  & \textbf{0} & 3 & 0 & 5 \\ \hline
		ACo & 496.98 & 1095 & \textbf{0} & 1 & 0 & 0 \\ \hline
		ERe & 160.64 & 704  & \textbf{0} & 2 & \textbf{52} & 100 \\ \hline
		STC & 547.14 & 1192 & 2 & 0 & 0  & 0 \\ \hline
		HE  & 187.13 & 910  & \textbf{0} & 2 & 13& 91 \\ \hline
		PCR & 112.72 & \textbf{481} & \textbf{0} & 2 & 46 & \textbf{104} \\ \hline
		iPCR& \textbf{103.73}   & 590 & \textbf{0} & \textbf{6} & 45 & 97 \\ \hline
	\end{tabular}  \label{tab:rank_er}
\end{table}

It can be seen from Table \ref{tab:rank_er} that iPCR receives the minimum average rank error $103.73$, followed by PCR. PCR obtains the minimum max rank error, followed by iPCR. Seven out of nine predictive measures receive a min rank error $0$, meaning that these measures predict at least one rank exactly as the true rank. iPCR predicts $6$ ranks correctly. ERe predicts $52$ top $10\%$ networks that are truly top $10\%$ networks, although their exact ranks may be different, which are followed by PCR and iPCR that correctly predict $46$ and $45$ networks respectively. Finally, PCR well predicts $104$ bottom $10\%$ networks, followed by ERe and iPCR. The test dataset contains $1200$ networks, hence there are $120$ networks ranked as top and bottom $10\%$, respectively.

It is thus clear that PCR and iPCR return better prediction measures than the spectral measures and the heterogeneity. More importantly, PCR and iPCR return not only the predictive measures, but also the entire controllability changing process of a network against the node-removal attack; while the spectral measures and heterogeneity return only a single quantitative value for the controllability robustness.

However, it is notable that PCR and iPCR require a substantial amount of training data, while the spectral measures and heterogeneity do not. Nevertheless, as demonstrated in \cite{Lou2020TCYB}, the overhead in training a CNN is quite low which, compared to the exhaustive attack simulation, is negligible.

\subsection{Computational Costs}
\label{sub:cost}

Compared to PCR, iPCR employs an extra CNN for classification, while the rest computation costs of PCR and iPCR are similar. Thus, the computational cost of iPCR is around two times of that of PCR. As discussed in \cite{Lou2020TCYB}, the cost of simulations to measure the controllability robustness is non-trivial. Compared to simulations, PCR accelerates the prediction speed by hundreds of times.

Given a PC with a 64-bit operation system, installed Intel i7-6700 (3.4 GHz) CPU, GeForce GTX 1080 Ti GPU, to collect a controllability curve for an ER network with $N=1000$ and $\langle k\rangle=5$ under random node attack, the elapsed time is about $162$ seconds for simulation, $0.42$ seconds for PCR, and $1.22$ seconds for iPCR. An example of run time comparison (including PCR, iPCR, and attack simulation) is available on web with source codes\footnote{\url{https://fylou.github.io/sourcecode.html}}.

\section{Conclusions}
\label{sec:end}

Network controllability robustness, which reflects how well a networked system can maintain its controllability after destructive attacks, is usually measured via attack simulations. Such an exhaustive simulation approach can return the true value of the controllability robustness, but is computationally costly and very time consuming. The predictor of controllability robustness (PCR) employs a single convolutional neural network (CNN) to successfully and efficiently achieve the prediction. In this paper, an improved multi-CNN and knowledge-based PCR (iPCR) is designed and evaluated, which takes advantage of prior knowledge from the given data. Extensive experimental studies, with thorough comparisons to seven other comparable measures, demonstrate that 1) iPCR predicts more precisely than PCR; 2) iPCR provides a better predictive measure than the traditional spectral measures and network heterogeneity.


\begin{thebibliography}{10}
	\providecommand{\url}[1]{#1}
	\csname url@samestyle\endcsname
	\providecommand{\newblock}{\relax}
	\providecommand{\bibinfo}[2]{#2}
	\providecommand{\BIBentrySTDinterwordspacing}{\spaceskip=0pt\relax}
	\providecommand{\BIBentryALTinterwordstretchfactor}{4}
	\providecommand{\BIBentryALTinterwordspacing}{\spaceskip=\fontdimen2\font plus
		\BIBentryALTinterwordstretchfactor\fontdimen3\font minus
		\fontdimen4\font\relax}
	\providecommand{\BIBforeignlanguage}[2]{{%
			\expandafter\ifx\csname l@#1\endcsname\relax
			\typeout{** WARNING: IEEEtran.bst: No hyphenation pattern has been}%
			\typeout{** loaded for the language `#1'. Using the pattern for}%
			\typeout{** the default language instead.}%
			\else
			\language=\csname l@#1\endcsname
			\fi
			#2}}
	\providecommand{\BIBdecl}{\relax}
	\BIBdecl
	
	\bibitem{Barabasi2016NS}
	A.-L. Barab{\'a}si, \emph{Network Science}.\hskip 1em plus 0.5em minus
	0.4em\relax Cambridge University Press, 2016.
	
	\bibitem{Newman2010N}
	M.~E. Newman, \emph{Networks: An Introduction}.\hskip 1em plus 0.5em minus
	0.4em\relax Oxford University Press, 2010.
	
	\bibitem{Chen2014Book}
	G.~Chen, X.~Wang, and X.~Li, \emph{Fundamentals of Complex Networks: Models,
		Structures and Dynamics}, 2nd~ed.\hskip 1em plus 0.5em minus 0.4em\relax John
	Wiley \& Sons, 2014.
	
	\bibitem{Chen2019Book}
	G.~Chen and Y.~Lou, \emph{Naming Game: {Models}, Simulations and
		Analysis}.\hskip 1em plus 0.5em minus 0.4em\relax Springer, 2019.
	
	\bibitem{Liu2011N}
	Y.-Y. Liu, J.-J. Slotine, and A.-L. Barab{\'a}si, ``Controllability of complex
	networks,'' \emph{Nature}, vol. 473, no. 7346, pp. 167--173, 2011.
	
	\bibitem{Yuan2013NC}
	Z.~Z. Yuan, C.~Zhao, Z.~R. Di, W.-X. Wang, and Y.-C. Lai, ``Exact
	controllability of complex networks,'' \emph{Nature Communications}, vol.~4,
	p. 2447, 2013.
	
	\bibitem{Posfai2013SR}
	M.~P{\'o}sfai, Y.-Y. Liu, J.-J. Slotine, and A.-L. Barab{\'a}si, ``Effect of
	correlations on network controllability,'' \emph{Scientific Reports}, vol.~3,
	p. 1067, 2013.
	
	\bibitem{Menichetti2014PRL}
	G.~Menichetti, L.~Dall'Asta, and G.~Bianconi, ``Network controllability is
	determined by the density of low in-degree and out-degree nodes,''
	\emph{Physical Review Letters}, vol. 113, no.~7, p. 078701, 2014.
	
	\bibitem{Motter15CHAOS}
	A.~E. Motter, ``Networkcontrology,'' \emph{Chaos: An Interdisciplinary Journal
		of Nonlinear Science}, vol.~25, no.~9, p. 097621, 2015.
	
	\bibitem{Wang2016AUTO}
	L.~Wang, X.~Wang, G.~Chen, and W.~K.~S. Tang, ``Controllability of networked
	mimo systems,'' \emph{Automatica}, vol.~69, pp. 405--409, 2016.
	
	\bibitem{Liu2016RMP}
	Y.-Y. Liu and A.-L. Barab{\'a}si, ``Control principles of complex systems,''
	\emph{Review of Modern Physics}, vol.~88, no.~3, p. 035006, 2016.
	
	\bibitem{Wang2017RSPTA}
	L.~Wang, X.~Wang, and G.~Chen, ``Controllability of networked
	higher-dimensional systems with one-dimensional communication channels,''
	\emph{Royal Society Philosophical Transactions A}, vol. 375, no. 2088, p.
	20160215, 2017.
	
	\bibitem{Wang2017SR}
	L.-Z. Wang, Y.-Z. Chen, W.-X. Wang, and Y.-C. Lai, ``Physical controllability
	of complex networks,'' \emph{Scientific Reports}, vol.~7, p. 40198, 2017.
	
	\bibitem{Zhang2017TAC}
	Y.~Zhang and T.~Zhou, ``Controllability analysis for a networked dynamic system
	with autonomous subsystems,'' \emph{IEEE Transactions on Automatic Control},
	vol.~62, no.~7, pp. 3408--3415, 2016.
	
	\bibitem{Xiang2019CSM}
	L.~Xiang, F.~Chen, W.~Ren, and G.~Chen, ``Advances in network
	controllability,'' \emph{IEEE Circuits and Systems Magazine}, vol.~19, no.~2,
	pp. 8--32, 2019.
	
	\bibitem{Liu2012PO}
	Y.-Y. Liu, J.-J. Slotine, and A.-L. Barab{\'a}si, ``Control centrality and
	hierarchical structure in complex networks,'' \emph{PLOS ONE}, vol.~7, no.~9,
	p. e44459, 2012.
	
	\bibitem{Wu2018JNS}
	E.~Wu-Yan, R.~F. Betzel, E.~Tang, S.~Gu, F.~Pasqualetti, and D.~S. Bassett,
	``Benchmarking measures of network controllability on canonical graph
	models,'' \emph{Journal of Nonlinear Science}, pp. 1--39, 2018.
	
	\bibitem{Zhang2019PA}
	R.~Zhang, X.~Wang, M.~Cheng, and T.~Jia, ``The evolution of network
	controllability in growing networks,'' \emph{Physica A: Statistical Mechanics
		and its Applications}, vol. 520, pp. 257--266, 2019.
	
	\bibitem{Hao2018IJRNC}
	Y.~Hao, Z.~Duan, and G.~Chen, ``Further on the controllability of networked
	{MIMO LTI} systems,'' \emph{International Journal of Robust and Nonlinear
		Control}, vol.~28, no.~5, pp. 1778--1788, 2018.
	
	\bibitem{Lou2018TCASI}
	Y.~Lou, L.~Wang, and G.~Chen, ``Toward stronger robustness of network
	controllability: {A} snapback network model,'' \emph{{IEEE} Transactions on
		Circuits and Systems {I}: {R}egular Papers}, vol.~65, no.~9, pp. 2983--2991,
	2018.
	
	\bibitem{Chen2019TCASII}
	G.~Chen, Y.~Lou, and L.~Wang, ``A comparative study on controllability
	robustness of complex networks,'' \emph{{IEEE} Transactions on Circuits and
		Systems {II}: {E}xpress Briefs}, vol.~66, no.~5, pp. 828--832, 2019.
	
	\bibitem{Lou2019R}
	Y.~Lou, L.~Wang, and G.~Chen, ``Enhancing controllability robustness of
	\textit{q}-snapback networks through redirecting edges,'' \emph{Research},
	vol. 2019, no. 7857534, 2019.
	
	\bibitem{Holme2002PRE}
	P.~Holme, B.~J. Kim, C.~N. Yoon, and S.~K. Han, ``Attack vulnerability of
	complex networks,'' \emph{Physical Review {E}}, vol.~65, no.~5, p. 056109,
	2002.
	
	\bibitem{Shargel2003PRL}
	B.~Shargel, H.~Sayama, I.~R. Epstein, and Y.~Bar-Yam, ``Optimization of
	robustness and connectivity in complex networks,'' \emph{Physical Review
		Letters}, vol.~90, no.~6, p. 068701, 2003.
	
	\bibitem{Schneider2011PNAS}
	C.~M. Schneider, A.~A. Moreira, J.~S. Andrade, S.~Havlin, and H.~J. Herrmann,
	``Mitigation of malicious attacks on networks,'' \emph{Proceedings of the
		National Academy of Sciences}, vol. 108, no.~10, pp. 3838--3841, 2011.
	
	\bibitem{Bashan2013NP}
	A.~Bashan, Y.~Berezin, S.~Buldyrev, and S.~Havlin, ``The extreme vulnerability
	of interdependent spatially embedded networks,'' \emph{Nature Physics},
	vol.~9, pp. 667--672, 2013.
	
	\bibitem{Xiao2014CPB}
	Y.-D. Xiao, S.-Y. Lao, L.-L. Hou, and L.~Bai, ``Optimization of robustness of
	network controllability against malicious attacks,'' \emph{Chinese Physics
		B}, vol.~23, no.~11, p. 118902, 2014.
	
	\bibitem{Wang2018TNSE}
	S.~Wang and J.~Liu, ``A multi-objective evolutionary algorithm for promoting
	the emergence of cooperation and controllable robustness on directed
	networks,'' \emph{IEEE Transactions on Network Science and Engineering},
	vol.~5, no.~2, pp. 92--100, 2018.
	
	\bibitem{Liang2015CPL}
	L.~Bai, Y.-D. Xiao, L.-L. Hou, and S.-Y. Lao, ``Smart rewiring: {I}mproving
	network robustness faster,'' \emph{Chinese Physics Letters}, vol.~32, no.~7,
	p. 078901, 2015.
	
	\bibitem{Chan2016DMKD}
	H.~Chan and L.~Akoglu, ``Optimizing network robustness by edge rewiring: {A}
	general framework,'' \emph{Data Mining and Knowledge Discovery}, vol.~30,
	no.~5, pp. 1395--1425, 2016.
	
	\bibitem{Yamashita2019COMPSAC}
	K.~Yamashita, Y.~Yasuda, R.~Nakamura, and H.~Ohsaki, ``On the predictability of
	network robustness from spectral measures,'' in \emph{2019 IEEE 43rd Annual
		Computer Software and Applications Conference (COMPSAC)}, vol.~2.\hskip 1em
	plus 0.5em minus 0.4em\relax IEEE, 2019, pp. 24--29.
	
	\bibitem{Hou2013ISDEA}
	L.~Hou, S.~Lao, B.~Jiang, and L.~Bai, ``Enhancing complex network
	controllability by rewiring links,'' in \emph{International Conference on
		Intelligent System Design and Engineering Applications ({ISDEA})}.\hskip 1em
	plus 0.5em minus 0.4em\relax IEEE, 2013, pp. 709--711.
	
	\bibitem{Xu2014CCDC}
	J.~Xu, J.~Wang, H.~Zhao, and S.~Jia, ``Improving controllability of complex
	networks by rewiring links regularly,'' in \emph{Chinese Control and Decision
		Conference ({CCDC})}, 2014, pp. 642--645.
	
	\bibitem{Liu2019ECCN}
	J.~Liu, H.~A. Abbass, and K.~C. Tan, ``Evolving robust networks using
	evolutionary algorithms,'' in \emph{Evolutionary Computation and Complex
		Networks}.\hskip 1em plus 0.5em minus 0.4em\relax Springer, 2019, pp.
	117--140.
	
	\bibitem{Wang2019IS}
	S.~Wang and J.~Liu, ``Designing comprehensively robust networks against
	intentional attacks and cascading failures,'' \emph{Information Sciences},
	vol. 478, pp. 125--140, 2019.
	
	\bibitem{Gunasekara2018MOO}
	R.~C. Gunasekara, C.~K. Mohan, and K.~Mehrotra, ``Multi-objective optimization
	to improve robustness in networks,'' in \emph{Multi-Objective
		Optimization}.\hskip 1em plus 0.5em minus 0.4em\relax Springer, 2018, pp.
	115--139.
	
	\bibitem{Zeng2012PRE}
	A.~Zeng and W.~Liu, ``Enhancing network robustness against malicious attacks,''
	\emph{Physical Review E}, vol.~85, no.~6, p. 066130, 2012.
	
	\bibitem{Wu2011PRE}
	Z.-X. Wu and P.~Holme, ``Onion structure and network robustness,''
	\emph{Physical Review E}, vol.~84, no.~2, p. 026106, 2011.
	
	\bibitem{Tanizawa2012PRE}
	T.~Tanizawa, S.~Havlin, and H.~E. Stanley, ``Robustness of onionlike correlated
	networks against targeted attacks,'' \emph{Physical Review E}, vol.~85,
	no.~4, p. 046109, 2012.
	
	\bibitem{Hayashi2018SR}
	Y.~Hayashi and N.~Uchiyama, ``Onion-like networks are both robust and
	resilient,'' \emph{Scientific Reports}, vol.~8, 2018.
	
	\bibitem{Ma2016PA}
	L.~Ma, J.~Liu, and B.~Duan, ``Evolution of network robustness under continuous
	topological changes,'' \emph{Physica A: Statistical Mechanics and its
		Applications}, vol. 451, pp. 623--631, 2016.
	
	\bibitem{Pu2012PA}
	C.-L. Pu, W.-J. Pei, and A.~Michaelson, ``Robustness analysis of network
	controllability,'' \emph{Physica {A}: {S}tatistical Mechanics and its
		Applications}, vol. 391, no.~18, pp. 4420--4425, 2012.
	
	\bibitem{Yan2016SR}
	X.-Y. Yan, W.-X. Wang, G.~Chen, and D.-H. Shi, ``Multiplex congruence network
	of natural numbers,'' \emph{Scientific Reports}, vol.~6, p. 23714, 2016.
	
	\bibitem{Lou2020TCASI}
	Y.~Lou, L.~Wang, K.-F. Tsang, and G.~Chen, ``Towards optimal robustness of
	network controllability: {A}n empirical necessary condition,'' \emph{IEEE
		Transactions on Circuits and Systems I: Regular Papers}, vol.~67, no.~9, pp.
	3163--3174, 2020.
	
	\bibitem{Schmidhuber2015NN}
	J.~Schmidhuber, ``Deep learning in neural networks: An overview,'' \emph{Neural
		Networks}, vol.~61, pp. 85--117, 2015.
	
	\bibitem{Wang2012ICPR}
	T.~Wang, D.~J. Wu, A.~Coates, and A.~Y. Ng, ``End-to-end text recognition with
	convolutional neural networks,'' in \emph{International Conference on Pattern
		Recognition (ICPR 2012)}.\hskip 1em plus 0.5em minus 0.4em\relax IEEE, 2012,
	pp. 3304--3308.
	
	\bibitem{Lai2015AAAI}
	S.~Lai, L.~Xu, K.~Liu, and J.~Zhao, ``Recurrent convolutional neural networks
	for text classification,'' in \emph{AAAI Conference on Artificial
		Intelligence}, 2015, pp. 2267--2273.
	
	\bibitem{Zhang2015NIPS}
	X.~Zhang, J.~Zhao, and Y.~LeCun, ``Character-level convolutional networks for
	text classification,'' in \emph{Advances in Neural Information Processing
		Systems (NIPS 2015)}, 2015, pp. 649--657.
	
	\bibitem{Li2015CVPR}
	H.~Li, Z.~Lin, X.~Shen, J.~Brandt, and G.~Hua, ``A convolutional neural network
	cascade for face detection,'' in \emph{IEEE Conference on Computer Vision and
		Pattern Recognition (CVPR)}, 2015, pp. 5325--5334.
	
	\bibitem{Ronneberger2015MICCAI}
	O.~Ronneberger, P.~Fischer, and T.~Brox, ``U-net: Convolutional networks for
	biomedical image segmentation,'' in \emph{International Conference on Medical
		Image Computing and Computer-Assisted Intervention}.\hskip 1em plus 0.5em
	minus 0.4em\relax Springer, 2015, pp. 234--241.
	
	\bibitem{Lou2020TCYB}
	Y.~Lou, Y.~He, L.~Wang, and G.~Chen, ``Predicting network controllability
	robustness: A convolutional neural network approach,'' \emph{IEEE
		Transactions on Cybernetics}, 2020, doi:10.1109/TCYB.2020.3013251.
	
	\bibitem{Ruths2013CNIV}
	J.~Ruths and D.~Ruths, ``Robustness of network controllability under edge
	removal,'' in \emph{Complex Networks IV}.\hskip 1em plus 0.5em minus
	0.4em\relax Springer, 2013, pp. 185--193.
	
	\bibitem{Simonyan2014arXiv}
	K.~Simonyan and A.~Zisserman, ``Very deep convolutional networks for
	large-scale image recognition,'' \emph{arXiv Preprint: 1409.1556}, 2014.
	
	\bibitem{Glorot2011ICAIS}
	X.~Glorot, A.~Bordes, and Y.~Bengio, ``Deep sparse rectifier neural networks,''
	in \emph{International Conference on Artificial Intelligence and Statistics},
	2011, pp. 315--323.
	
	\bibitem{Bishop2006}
	C.~M. Bishop, \emph{Pattern Recognition and Machine Learning}.\hskip 1em plus
	0.5em minus 0.4em\relax springer, 2006.
	
	\bibitem{Niepert2016ICML}
	M.~Niepert, M.~Ahmed, and K.~Kutzkov, ``Learning convolutional neural networks
	for graphs,'' in \emph{International Conference on Machine Learning (ICML)},
	2016, pp. 2014--2023.
	
	\bibitem{Barabasi1999SCI}
	A.-L. Barab{\'a}si and R.~Albert, ``Emergence of scaling in random networks,''
	\emph{Science}, vol. 286, no. 5439, pp. 509--512, 1999.
	
	\bibitem{Erdos1964RG}
	P.~Erd{\"{o}}s and A.~R{\'e}nyi, ``On the strength of connectedness of a random
	graph,'' \emph{Acta Mathematica Hungarica}, vol.~12, no. 1--2, pp. 261--267,
	1964.
	
	\bibitem{Newman1999PLA}
	M.~E. Newman and D.~J. Watts, ``Renormalization group analysis of the
	small-world network model,'' \emph{Physics Letters A}, vol. 263, no. 4--6,
	pp. 341--346, 1999.
	
\end{thebibliography}
\end{document}